# Title: Correlated insulator of excitons in WSe$_2$/WS$_2$ moiré superlattices

**Authors:** Richen Xiong[1], Jacob H. Nie[1], Samuel L. Brantly[1], Patrick Hays[2], Renee Sailus[2], Kenji Watanabe[3], Takashi Taniguchi[4], Sefaattin Tongay[2], Chenhao Jin[1]*

**Affiliations:**
[1]*Department of Physics, University of California at Santa Barbara; Santa Barbara, CA 93106, USA*
[2]*School for Engineering of Matter, Transport, and Energy, Arizona State University; Tempe 85287 AZ, USA.*
[3]*Research Center for Functional Materials, National Institute for Materials Science; 1-1 Namiki, Tsukuba 305-0044, Japan.*
[4]*International Center for Materials Nanoarchitectonics, National Institute for Materials Science; 1-1 Namiki, Tsukuba 305-0044, Japan*

* Corresponding author. Email: jinchenhao@ucsb.edu

**Abstract:** A panoply of unconventional electronic states has been observed in moiré superlattices. Engineering similar bosonic phases remains, however, largely unexplored. We report the observation of a bosonic correlated insulator in WSe$_2$/WS$_2$ moiré superlattices composed of excitons, i.e., tightly bound electron-hole pairs. We develop a pump probe spectroscopy method that we use to observe an exciton incompressible state at exciton filling $v_{ex}$ = 1 and charge neutrality, indicating a correlated insulator of excitons. With varying charge density, the bosonic correlated insulator continuously transitions into an electron correlated insulator at charge filling $v_e$ = 1, suggesting a mixed correlated insulating state between the two limits. Our studies establish semiconducting moiré superlattices as an intriguing platform for engineering bosonic phases.

**Main Text:**

Strongly correlated phases can emerge in flat-band systems when many-body interactions dominate over kinetic energy (*1, 2*). Semiconducting transition metal dichalcogenide (TMDC) moiré superlattices offer a unique platform where both fermionic and bosonic quasi-particles – charges(*3–12*) and excitons(*13–21*) – occupy flat-bands. Previous studies have primarily focused on the fermionic sector with exciton filling $v_{ex} = 0$, such as Mott and Wigner crystal states(*3–8*), stripe phase(*9*), continuous Mott transition(*10, 11*), quantum anomalous Hall insulators(*12*), along with single exciton behavior in the $v_{ex} \to 0$ limit, such as moiré excitons(*13–16*) and trions(*20, 21*). Recently, increasing efforts have been put into an intermediate exciton density regime of $v_{ex} \sim 0.1$. Examples include studies of exciton-mediated ferromagnetism(*22*) as well as excitonic insulators(*23, 24*) -- insulators for charge but "metals" for excitons, where electron behavior is affected by the coexisting excitons. However, in these prior studies, excitons themselves are in a compressible fluid state; strongly correlated phases of bosons remain elusive.

Here we explore the high exciton density regime of $v_{ex} \sim 1$. We create and identify a bosonic correlated insulator consisting of interlayer excitons in a 60-degree-aligned $WSe_2/WS_2$ moiré superlattice. Each interlayer exciton contains an electron in the $WS_2$ layer and a hole in $WSe_2$ with a large binding energy of hundreds of meV(*25*), which is the ground state exciton configuration of a type II heterostructure (Fig. 1A). We find an exciton incompressible state at $v_{ex} = 1$, i.e., one exciton per moiré site, a hallmark of a bosonic correlated insulator. We further study the phase diagram spanned by $v_e$ and $v_{ex}$ and observe a mixed correlated insulator along the path of $v_{tot} = v_{ex} + v_e = 1$. Owing to the large exciton binding energy and strong correlations in this system, the bosonic correlated insulator persists to above 30 K, orders of magnitude higher than previous studies in cold atoms and quantum wells(26, 27). Our results highlight semiconducting moiré superlattices as an attractive platform for engineering novel bosonic phases at high temperature, such as valley pseudospin order and pseudospin liquid, bosonic Mott-superfluidity transition(28), exciton-mediated superconductivity(29) and topological excitons(18, 19), as well as for exploring the many-body physics of interacting fermions and bosons.

**Direct measurement of exciton compressibility**

One major challenge in identifying a correlated insulator of excitons is to distinguish it from disordered excitons without a lattice. Several works have reported power-dependent PL spectra in TMDC systems(30, 31) and emergence of additional high energy peaks under strong excitation. However, this only indicates the existence of local exciton-exciton interactions while providing no information on an ordered exciton lattice. The key evidence of a correlated insulator, as widely recognized in electrical measurements, is an incompressible state at particular lattice fillings(32). Optical measurements such as PL, on the other hand, typically collect responses from all excitons in the system and cannot obtain compressibility information. Here we develop a pump-probe spectroscopy method that directly measures exciton compressibility (Fig. 1B); this is an optical analogue of electrical capacitance measurements(33) and distinct from conventional optical pump-probe configurations. In an electronic capacitance measurement, a DC "pump" gate voltage tunes the background charge density and a small AC

"probe" voltage slightly modulates the charge density. Similarly, here a relatively strong pump light tunes the background exciton density, and the weak probe light injects a small number of additional excitons and detects their response. In both cases, the charge/exciton compressibility can be directly obtained from the minimum energy it takes to add one more particle on top of a given background particle density. Importantly, a DC "pump" is necessary to maintain a stable background particle density and a well-defined ground state, whereas the "probe" needs to be AC modulated to isolate the responses of particles created by the probe(*34*).

**Correlated insulator of bosons**

With the capability of tuning charge and exciton density through electrostatic gating and pump light respectively, we fully explore the phase diagram spanned by the charge filling $v_e$ and exciton filling $v_{ex}$. We start with the axes, i.e., $v_{ex} = 0$ or $v_e = 0$. Figures 1, C and E, show the PL and absorption spectra, respectively, of a 60-degree-aligned moiré bilayer (device D1) at $v_{ex} = 0$ (zero pump intensity) and $v_e \geq 0$ (see Fig. S2 for complete doping dependence). At charge neutrality, the PL features a single peak at 1.43 eV from interlayer exciton emission(*6, 8*), whereas the absorption shows three peaks from moiré intralayer excitons(*13*). At approximately $v_e = 1$ and 2 ($v_e = 1$ corresponds to the moiré density $n_0 = 2.1 \times 10^{12}$ cm$^{-2}$ (*34*)), the emission peak blueshifts suddenly, and the absorption peaks show a kink. These features originate from the emergence of insulating states at integer electron fillings, which has been independently confirmed by capacitance and microwave impedance measurements(*3, 6*). The PL energy jump at $v_e = 1$ can be intuitively understood from the emergence of a fermionic correlated insulator state where all available sites are occupied by one electron; any additional excitons injected are therefore forced into a higher energy state. We will discuss the mechanisms of these spectral changes in more depth in the discussion section.

We now turn to the case of $v_e = 0$ by fixing the gate voltage $V_g$ at charge-neutral -0.5 V. Figures 1, D and F, show the dependence of pump-probe PL and absorption spectra on the pump light intensity, which effectively controls $v_{ex}$. To account for the non-linear dependence of $v_{ex}$ on pump intensity, we also show on the right axes dipolar-interaction-induced interlayer exciton energy shift $\Delta_{dipole}$ (the energy shift of peak I relative to zero pump intensity, see Fig. 1D), which is approximately proportional to $v_{ex}$(*34–36*). Interestingly, a jump in the PL energy is observed here as well. This jump is well reproduced in two other 60-degree-aligned moiré bilayers and one 0-degree aligned moiré bilayer but is absent in a slightly misaligned bilayer(*34*), indicating that its origin is from correlation effects. The qualitative similarity to the gate-dependent PL suggests a similar origin behind the exciton energy jump: the emergence of a particle lattice that occupies all available sites. The most natural candidates are interlayer excitons, which are the immediate products of pump light absorption, and interlayer charge transfer in a type II heterojunction. To elucidate the nature of the lattice, we compare its effects on PL and absorption spectra to those from an electron lattice. PL spectra probe interlayer exciton responses. Although both the gate-induced electron lattice (Fig. 1C) and pump-induced lattice (Fig. 1D) lead to a jump in interlayer exciton energy in PL, the amplitudes are quite different: 35 meV and 15 meV, respectively. A more prominent difference is observed in the absorption spectra, which examines how intralayer excitons respond to the induced lattices. Electrons induce rich features, such as shifting, merging and splitting of exciton resonances (Fig. 1E). In contrast, the pump-injected particles have rather weak effects on intralayer excitons, only slightly decreasing their oscillator strength (Fig. 1F). These distinctive behaviors indicate that the pump-induced lattice is not formed by electrons.

To further confirm the exciton nature of the pump-induced lattice, we move away from the axes and investigate the phase diagram at $v_{ex} > 0$ and $v_e > 0$. Fig. 2A shows a set of pump intensity-dependent PL spectra at different gate voltages, corresponding to increasing electron density in the $WS_2$ layer. All plots have qualitatively similar behaviors – interlayer exciton energy shows a jump. Upon increasing $v_e$, the jump occurs earlier and earlier, eventually appearing at zero pump intensity when $v_e = 1$ ($V_g = 0.8$ V), consistent with the fact that gate-injected electrons have already occupied all sites at $v_e = 1$. Based on these observations, the pump-injected particles will occupy sites previously available to free electrons in $WS_2$ and form a mixed lattice together with free electrons. Besides an electron itself, which has been already excluded, the only other candidate for the pump-injected particle is the interlayer exciton composed of an electron in the $WS_2$ layer and a hole in the $WSe_2$ layer.

We, therefore, conclude that the pump light is creating interlayer excitons and tuning $v_{ex}$. At charge neutrality, the observed jump in PL spectra then corresponds to a sudden increase in the interlayer exciton energy when increasing its density, i.e., an incompressible state of the interlayer exciton. Moreover, this state connects smoothly into the $v_e = 1$ electron correlated insulator, indicating one particle per moiré site along the entire path and $v_{ex} = 1$ in the limit of sole occupation by excitons (we have also independently calibrated the exciton density, see ref. (*34*)). Our observation of an incompressible exciton state at $v_{ex} = 1$ (half-filling of an exciton band considering the valley degeneracy) and charge neutrality is a hallmark of a correlated insulator purely made of excitons. This correlated insulator could be a bosonic Mott insulator (*28*, *37*), a generalized Wigner crystal(*38*, *39*) or a charge transfer insulator(*40*). We unambiguously exclude the charge transfer insulator scenario through valley (flavor)-resolved measurements(*34*). Estimations of $r_s$ (ratio of Coulomb interaction to kinetic energy) and the Mott criterion suggest a bosonic Mott insulator nature of our observation (*34*, *38*, *39*).

**Mixed correlated insulator**

To quantify the jump in interlayer exciton energy, we fit PL spectra with two Lorentzian peaks(*34*) and compute the exciton energy change $\Delta E_{ex} = [(E_1 I_1 + E_2 I_2)/(I_1 + I_2) - E_1]$, which can be considered as an effective exciton chemical potential. Here $E_1$ ($E_2$) and $I_1$ ($I_2$) correspond to the energy and amplitude of the PL peak I and II, respectively (see Fig. 1D). Figure 2C summarizes the evolution of $\Delta E_{ex}$ with pump intensity at representative gating, where a transition in exciton energy is clearly observed at all gate voltages. The transition appears not particularly sharp around charge neutrality, largely because of the non-linear dependence of exciton density on pump intensity (Fig. 1D). In addition, the transition is broadened by spatial inhomogeneity in the exciton density that is expected to be much larger than the charge case (*34*). At charge neutrality $V_g$ = -0.5 V, we determine the position of $v_{ex}$ = 1 from the middle point of the $\Delta E_{ex}$ transition. We also independently determine the exciton density at this point to be $(2\pm0.2)\times10^{12}$ cm$^{-2}$ from time-resolved measurements (*34*), which matches well with the expected density of moiré cells $n_0 = 2.1\times10^{12}$ cm$^{-2}$. At finite electron density, the transition happens when electrons and excitons cooperate to occupy all available sites. The transition points therefore correspond to $v_{tot} = v_e + v_{ex} = 1$ (green triangles in Fig. 2C) and keep shifting to lower pump intensity until reaching $v_{ex} = 0$ when $v_e = 1$.

To separate $v_e$ and $v_{ex}$ better, we also measure absorption spectra of the moiré superlattice while varying gate voltage and pump intensity. Because intralayer excitons are only sensitive to $v_e$ but not $v_{ex}$ (Fig. 1, E and F), we use spectral changes in absorption to independently determine $v_e$. Figure 2B shows the gate-dependent absorption spectra of the moiré bilayer at different pump intensities. All plots show similar behaviors except for a slight shift in gate voltage. This can be seen more clearly in the gate-differentiated absorption spectra at representative pump intensity (Fig. 2D) (*34*). Blue triangles in Fig. 2, B and D, label the position of "kinks" that correspond to $v_e = 1$.

With both $v_{tot}$ and $v_e$ extracted, Fig. 3A summarizes the phase diagram spanned by $v_e$ and $v_{ex}$ at base temperature of 1.65 K. The color scale represents $\Delta E_{ex}$; the boundaries of $v_{tot} = 1$ and $v_e = 1$ are overlaid on the phase diagrams. The $v_e = 1$ line shifts slightly towards lower gate voltage at high pump intensity, presumably from photocarriers generated during the relaxation of excitons. However, this effect is rather weak, and the $v_e = 1$ line deviates far from the $v_{tot} = 1$ line where transitions in $\Delta E_{ex}$ are observed. This further confirms that the transition at $v_{tot} = 1$ is not from an electron lattice. Instead, it originates from a mixed lattice composed of both excitons and electrons. Our observations, therefore, suggest a mixed correlated insulator along the $v_{tot} = 1$ line, which smoothly connects into a bosonic (fermionic) correlated insulator at the end point of $v_{ex} = 1$ ($v_e = 1$).

We further obtain the phase diagrams at 15 and 30 K (Fig. 3, B and C, see Fig. S9-S12 for details). A sharp transition in $\Delta E_{ex}$ is always observed at $v_e = 1$, consistent with the strong charge correlation and a high melting temperature of charge correlated insulator in WSe$_2$/WS$_2$ moiré superlattices(*3, 4*). Similarly, the behavior at $v_{ex} = 1$ remains largely unchanged with temperature, indicating survival of the bosonic correlated insulator to above 30 K. In contrast, for the regions in between, the transition in $\Delta E_{ex}$ becomes much slower than at base temperature. This can be seen clearly by comparing the $\Delta E_{ex}$ evolution at charge neutral ($V_g = -0.5$ V) and finite electron density ($V_g = -0.1$ V), as shown in Fig. 3, D-F. At base temperature the finite $v_e$ case shows a sharper rise than charge neutral (Fig. 3D), which can be naturally explained by the smaller $v_{ex}$ needed to reach $v_{tot} = 1$ and therefore less inhomogeneous broadening from exciton density. Interestingly, at 30 K the rise in $\Delta E_{ex}$ becomes smoother at $V_g = -0.1$ V compared to charge neutral, indicating increasing exciton compressibility and partial melting of the mixed correlated insulator. This observation suggests that the mixed correlated insulator is less stable than both components(*34*).

**Discussion**

To allow direct comparison with theory, we calibrate $v_{ex}$ throughout the phase diagram by time-resolved measurements(*34*), as shown in Fig. 4A. The $\Delta E_{ex}$ jumps (green triangles) indeed occur at $v_{ex} + v_e = 1$ (green dashed line), further supporting the mixed correlated insulator origin. The experimental phase diagram matches very well with predictions from a two-component Hubbard model with both fermionic and bosonic species(*34*). Along the horizontal axis ($v_{ex} = 0$), $\Delta E_{ex}$ shows a sudden jump at $v_e = 1$. This can be understood as excitons always avoid electron-occupied sites owing to the on-site electron-exciton repulsion (Fig. 4B, see discussions in (*34*)) until all sites are electron-occupied at $v_e = 1$, after which adding an exciton necessarily pays the additional energy cost of $U_{e-ex}$ (Fig. 4D). We thereby determine $U_{e-ex}$ to be 32~40 meV (varies between samples). Such doping dependence is distinctively different from that in monolayer TMDC or TMDC bilayers where strong correlation is not observed, such as WSe$_2$/MoSe$_2$(*20, 21*). In these systems, electron doping immediately leads to the formation of trion and a corresponding emission peak in PL at lower energy(*20, 21*). Here in WSe$_2$/WS$_2$, in contrast, the emission spectrum remains largely unchanged until $v_e = 1$, after which the PL shows a blueshift of ~35 meV and ~20 meV in 60- and 0- twisted moiré superlattices, respectively (*6, 8*). Our model provides a natural explanation to this widely observed yet not fully understood PL doping dependence.

Similarly, $\Delta E_{ex}$ remains 0 along the vertical axis until $v_{ex} = 1$, after which adding an exciton requires an additional energy cost of $U_{ex-ex}$ (Fig. 4C). We thereby determine $U_{ex-ex}$ to be 14~20 meV (varies between samples). Away from the axes, $\Delta E_{ex}$ remains 0 for $v_{ex} + v_e < 1$ (region I in Fig. 4A) because added excitons can find an empty site to avoid on-site repulsion. For $v_{ex} + v_e \geq 1$ but $v_e < 1$ (region II), added excitons will prefer to stay on an exciton site because $U_{ex-ex} < U_{e-ex}$, and the additional energy cost is $\Delta E_{ex} = U_{ex-ex}$. For $v_e \geq 1$ (region III), all sites are already occupied by electrons and added excitons can only reside on an electron site; therefore $\Delta E_{ex} = U_{e-ex}$.

The above analysis ignores the hopping term or the two pseudospins of excitons from the *K* and *K'* valleys(*25*). Nevertheless, it captures all salient features of the experiments. Several interesting theoretical questions remain to fully understand the phase diagram, such as on the effects of finite hopping, the microscopic mechanism of exciton-exciton interaction and its dependence on exciton pseudospin, for which our results provide a valuable experimental reference.

**Acknowledgement:** R. X. thanks T. Xie, C. Tschirhart, A. Potts, Y. Guo, W. Gong, W. Y. Xing for experimental help. **Funding:** C.J. acknowledges support from Air Force Office of Scientific Research under award FA9550-23-1-0117. R.X. acknowledges support from the UC Santa Barbara NSF Quantum Foundry funded via the Q-AMASE-i program under award DMR-1906325. S.T. acknowledges support from DOE-SC0020653 (materials synthesis), NSF DMR-2206987 (elemental purification), NSF ECCS 2052527 (electrical characterization for crystal optimization), DMR 2111812 (excitonic characterization during growth optimization), and CMMI 2129412 (large size crystal development). K.W. and T.T. acknowledge support from the JSPS KAKENHI (Grant Numbers 19H05790, 20H00354 and 21H05233).

**Author contributions:** C.J. conceived and supervised the project. R.X. and J.H.N. fabricated the devices. R.X. and S.L.B performed the optical measurements and analyzed the data. P.H., R.S. and S.T. grew the WSe$_2$ and WS$_2$ crystals. K.W. and T.T. grew the hBN crystals. C.J. and R.X. wrote the manuscript with the input from all the authors.

**Competing interests:** The authors declare no competing interests.

**Data and materials availability:** All data in the main text and supplementary materials, as well as the code for peak fitting, are available from Open Science Framework(*41*).

**Supplementary Materials：**

Materials and Methods

Supplementary Text

Figs. S1 to S16

References (*42*–*63*)

# Figure 1

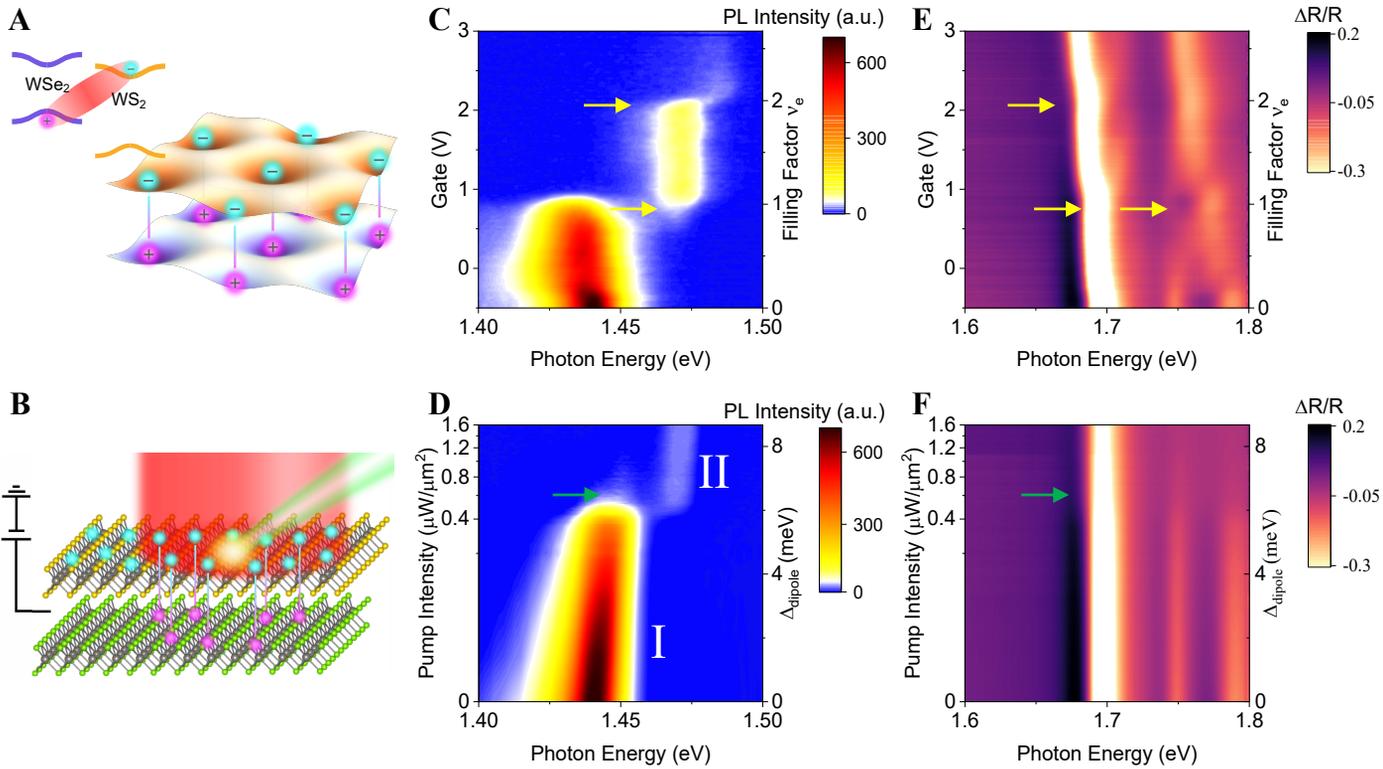

**Fig. 1. Bosonic correlated insulator. (A)** Illustration of a bosonic correlated insulator consisting of interlayer excitons. Magenta spheres indicate holes and cyan spheres electrons. Inset: type II band alignment of $WSe_2/WS_2$ heterostructure. **(B)** Schematics of continuous wave pump probe spectroscopy. The exciton and electron density are independently controlled by pump light and electrostatic gate. Red and green shading correspond to wide-field pump light and focused probe light, respectively. **(C,E)** Gate-dependent PL (C) and absorption (E) spectra of a 60-degree aligned $WSe_2/WS_2$ moiré bilayer (device D1) at zero pump intensity. The PL peak shows a sudden blue shift at electron filling $v_e = 1$ and 2 (yellow arrows), where absorption spectrum shows kinks and splitting. **(D,F)** Pump intensity-dependent PL (D) and absorption (F) spectra of device D1 at charge neutrality. Right axes show dipolar-interaction-induced interlayer exciton energy shift $\Delta_{dipole}$, which is approximately proportional to $v_{ex}$. The dominant PL peak in (D) at low and high pump intensity are labeled as peak I and II, respectively. All measurements are performed at base temperature of 1.65 K.

Figure 2

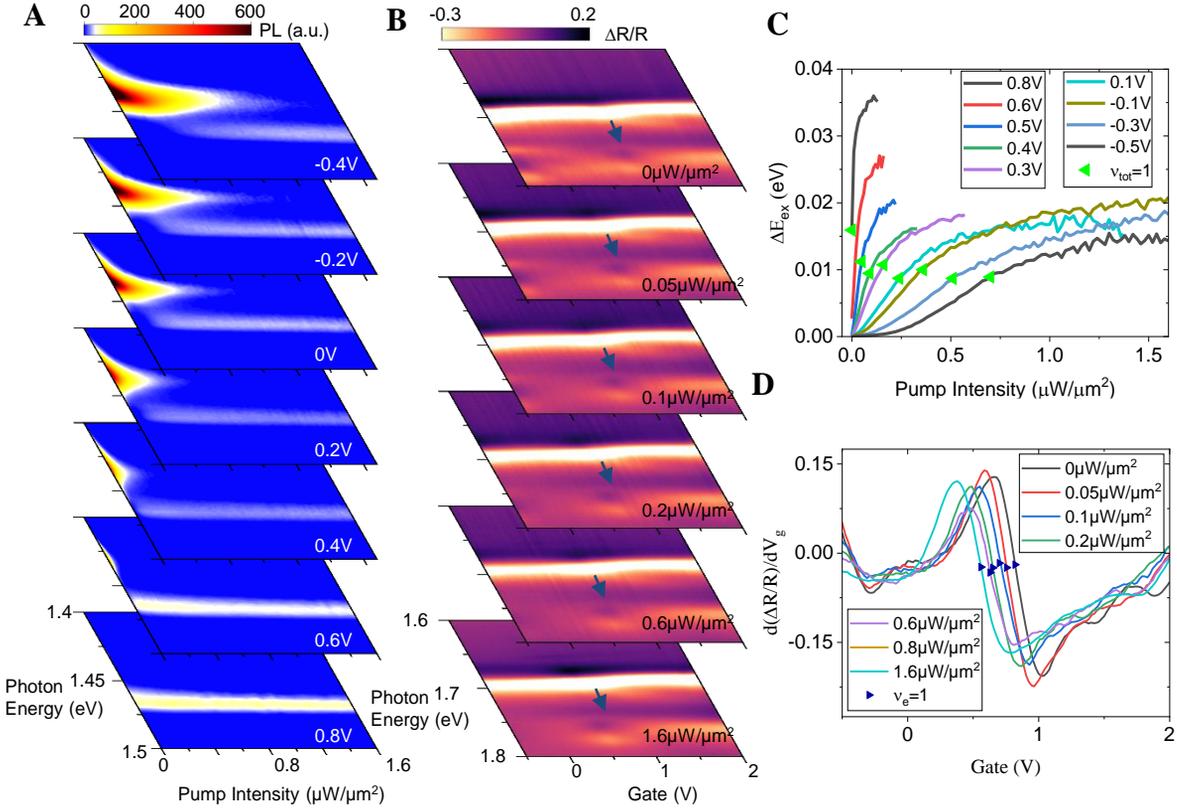

**Fig. 2. Mixed correlated insulator.** (A) Pump intensity-dependent PL spectra at gate voltages $V_g$ from -0.4 V (near charge neutrality) to 0.8 V (electron-one-filling). (B) Gate-dependent absorption spectra at pump intensities from 0 to 1.6 μW/μm². Blue arrows mark the kink and splitting in absorption peaks at $v_e = 1$. (C) Power dependent interlayer exciton energy change $\Delta E_{ex}$ at representative gate voltages. Green triangles mark middles of the transitions, which appear at smaller pump intensity ($v_{ex}$) with increasing gate voltage ($v_e$), consistent with a mixed correlated insulator state at $v_{tot} = 1$. (D) First order derivative of absorption spectra with respect to gate voltage at 1.76 eV under different pump intensities. Blue triangles denote $v_e = 1$. All measurements are performed at base temperature of 1.65 K in device D1.

Figure 3

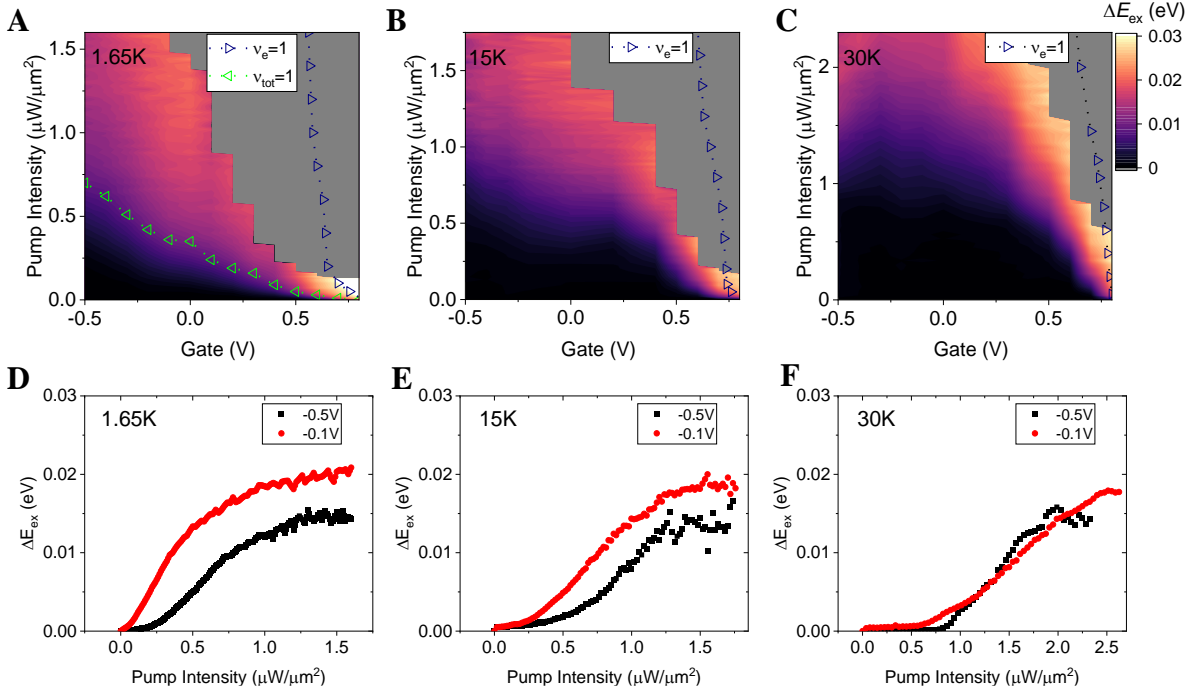

**Fig. 3. Phase diagram.** **(A-C)** $\Delta E_{ex}$ with respect to the gate voltage and pump intensity at 1.65 K (A), 15 K (B) and 30 K (C) measured on device D1. The boundaries of $v_{tot} = 1$ and $v_e = 1$, as determined from the pump-probe PL and absorption measurements, respectively, are highlighted with green and blue triangles. The clear separation between the two boundaries confirms that the transition of $\Delta E_{ex}$ at $v_{tot} = 1$ is not from a charge correlated insulator state until close to $v_e = 1$. Regions of high pump intensity and/or high gating are not shown because peak I has already disappeared and cannot be fitted reliably(34). **(D-F)** Vertical line-cuts of the phase diagrams for $V_g = -0.5$ V (charge neutrality) and -0.1 V (electron doped) at 1.65 K (D), 15 K (E) and 30 K (F). Whereas at 1.65 K $\Delta E_{ex}$ rises faster for $V_g = -0.1$ V than -0.5 V, at 30 K it rises slower for $V_g = -0.1$ V than -0.5 V, suggesting partial melting and lower stability of the mixed correlated insulator state.

Figure 4

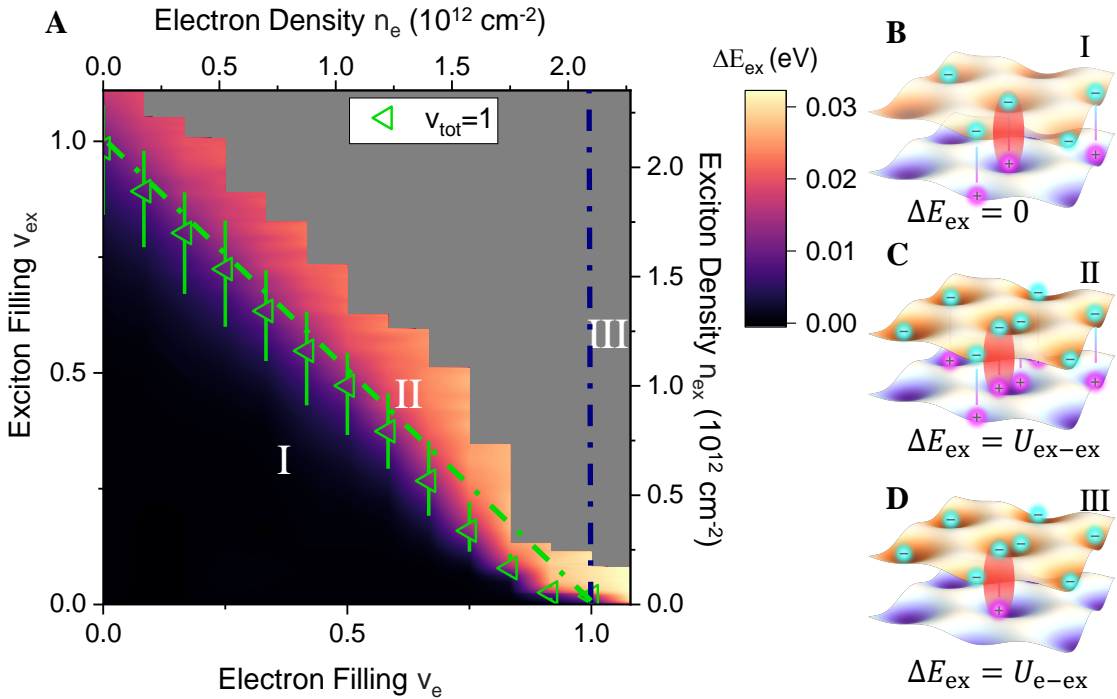

**Fig. 4. Mixed Hubbard model. (A)** $\Delta E_{ex}$ phase diagram measured on another 60-degree-aligned moiré bilayer (device D2) at 1.65 K with calibrated $v_{ex}$ and $v_e$. Green ($v_{tot} = 1$) and blue ($v_e = 1$) dashed lines are boundaries predicted by a mixed Hubbard model, where $\Delta E_{ex}$ is expected to jump. Green triangles label the experimental 50% transition point in $\Delta E_{ex}$, which matches well with the prediction. Error bars denote the range between 20% to 80% transition in $\Delta E_{ex}$. **(B-D)** Schematics showing the energy required to add one interlayer exciton (red) into the system for region I (B), II (C) and III (D) in the phase diagram. Additional energy cost of $\Delta E_{ex}=U_{ex\text{-}ex}$ and $U_{e\text{-}ex}$ is required in region II and III, respectively, from on-site repulsion between excitons and between electron-exciton.

# Supplementary Materials for

## Correlated insulator of excitons in WSe$_2$/WS$_2$ moiré superlattices

Richen Xiong, Jacob H. Nie, Samuel L. Brantly, Patrick Hays, Renee Sailus, Kenji Watanabe, Takashi Taniguchi, Sefaattin Tongay, Chenhao Jin

Correspondence to: jinchenhao@ucsb.edu

**This PDF file includes:**

Materials and Methods

Supplementary Text

Figs. S1 to S16

References

**Contents**

**Materials and Methods**

1. Device fabrication and characterization: Monolayer $WS_2$, monolayer $WSe_2$, few-layer graphite and thin hexagonal boron nitride (hBN) flakes were exfoliated onto silicon substrates with a 285 nm silicon oxide layer. Polarization-resolved second harmonic generation (SHG) was used to determine the relative angles between the $WSe_2$ and $WS_2$ crystalline axes (see Fig. S1B). The hBN flakes with a thickness of around 20 nm were used as the gate dielectrics. Few-layer graphite flakes were used as the contact electrode and gates. The $WSe_2/WS_2$ heterostructures with symmetric top and bottom gates were built using a layer-by-layer dry transfer method (*42*). The whole sample was then released to a 90nm $Si/SiO_2$ substrate. The stamps for picking up flakes were made of a film of polycarbonate on polydimethylsiloxane. The polycarbonate residue on the sample was removed by dissolving in chloroform, followed by a rinse in isopropyl alcohol. Contacts (~100 nm gold with ~5 nm chromium and ~15nm palladium adhesion layers) to the few-layer graphite were made by electron-beam lithography, plasma etching and electron-beam evaporation. Fig. S1A shows an optical image of a typical device. Since the top and bottom gates are nearly symmetric, we applied the same gate voltage $V_g$ to top and bottom gates to tune carrier concentration of the sample while keeping the out-of-plane electric field near zero. The sample was electron-doped with positive gate voltage.

2. Continuous-wave pump-probe spectroscopy: The samples were mounted in a closed-cycle cryostat (Quantum Design, OptiCool). All optical measurements were performed at base temperature of 1.65 K unless otherwise specified. A continuous wave 660 nm diode laser was used as the pump light with beam size of around 100 $\mu m^2$. The large pump beam size ensures a homogeneous intensity in the center region that is inspected by the probe beam. A 532 nm green laser and a broadband tungsten lamp were used as the probe light for PL and absorption measurements respectively. The beam size of probe light was around 4 $\mu m^2$. The probe power intensity was kept below 10 nW/$\mu m^2$, while the pump intensity ranged from 0 to 3 $\mu W/\mu m^2$. To isolate the response from probe-created excitons, the probe light was modulated by an optical chopper at frequency of 8 Hz. The signal was detected by a liquid-nitrogen-cooled CCD camera coupled with a spectrometer, which was externally triggered at 16 Hz and phase locked to the chopper. The spectra with and without the probe light were thereby obtained, and their difference gives the signals only from the probe light (see Fig. 1D for an example). To help isolate the probe light response, we also implement a spatial filter at a conjugate image plane of the sample, which only allows light from the probe-covered region to go through. The reflection contrast was extracted as $\Delta R/R = (R' - R)/R$, where $R'$ and $R$ are the reflected light spectrum and the featureless reference spectrum when the sample is highly hole-doped, respectively.

**Supplementary Text**

3. Pump-probe PL on the hole doping side: We find that hole doping does not cooperate with excitons to form a mixed correlated insulator, and instead will destabilize the correlated insulator (see Fig. S3A, B for examples). This can be potentially understood from that the lowest energy conduction band orbital (CBM) and highest energy valence band orbital (VBM) are localized around different high symmetry points within a moiré unit cell, as recently reported by both theory and experiments(*43–46*). Our experiment on the electron side indicates that exciton and CBM electron are likely localized around the same high symmetry points, which is then different from the VBM orbital. As a result, adding holes will not cooperate with exciton in forming a lattice, and may act as effective disorders to destabilize the lattice.

4. Sample quality and alignment dependence: The PL linewidth of device D1 and D2 are 15 and 12 meV at charge neutrality, respectively, which is comparable to the best value in literature. This indicates the high quality of our devices. To further test the effects of sample quality, we have performed measurements on another device D3. It has broader PL linewidth of ~ 20 meV, indicating lower sample quality. We find an $U_{\text{e-ex}}$ of 32 meV (Fig. S5A) and $U_{\text{ex-ex}}$ of 14 meV (Fig. S5B), consistent with device D1 and D2. This indicates that $U_{\text{e-ex}}$ and $U_{\text{ex-ex}}$ are quite robust against disorders, as expected from a large correlation strength.

On the other hand, emergence of bosonic and mixed correlated insulator requires flat exciton bands and strong correlation, which depends critically on the layer alignment. Our observations are well-reproduced across all devices with twist angle error <~1 degree from 60-degree. The good reproducibility benefits from the 4% intrinsic lattice mismatch between $WSe_2/WS_2$, which makes the moiré periodicity and effects insensitive to a small twist angle and remain virtually intact for twist angle error <~1 degree. Whereas, for twist angle error >~2 degrees, the moiré periodicity and lattice reconstruction start to change considerably, which in turn strongly affects the moiré potential, bandwidth, wavefunction and correlation strength such as $U_{\text{e-ex}}$ and $U_{\text{ex-ex}}$. For example, moiré excitons disappear for twist angle error >~3 degrees due to the sharp decrease of moiré potential(*13, 47*). For twist angle error >~5 degrees, the two layers are effectively decoupled and will recover monolayer properties(*47, 48*). We have measured a 3-degree misaligned sample D4 as a reference, where no bosonic or mixed correlated insulator state is observed (Fig. S6). This is consistent with a dramatic decrease of moiré potential and correlation strength at twist angle error >~2 degrees.

We have also performed pump-probe PL measurements of the 0-degree device D5 and observed similar responses from bosonic correlated insulator at charge neutrality (Fig. S7B). Under both electron- and hole-doping, the bosonic correlated insulator is destabilized and disappeared under moderate doping (Fig. S7 C&D), similar to the hole-doped case of 60-degree devices. The two different types of doping behaviors, cooperation or destabilization, likely depend on whether charges and excitons are localized around the same high symmetry point within a moiré unit cell. Investigation on this topic would be of interest for future studies.

5. Estimation of $r_s$ and Mott criterion: We estimate $r_s$ and Mott criterion to distinguish between a bosonic Mott insulator and a generalized Wigner crystal (GWC) state. The estimation of Mott criterion in the exciton case is fundamentally different from the charge case: electron crystallization can be intuitively understood from a competition between the kinetic energy $W$ and the potential energy $U$. While the kinetic energy of excitons can be estimated in a similar manner as electrons, the long-range interaction of excitons is through dipole-dipole interaction and scales as $1/r^3$, instead of $1/r$ for electrons. As a result, the potential energy of excitons is much weaker than the electron case, resulting in a much smaller $r_s \sim U/W$ (*38, 39*) and less favored GWC state. Quantitatively, $U$ of interlayer excitons can be estimated as $p_{\text{ex}}^2/(4\pi\epsilon\lambda_m^3)$, where $p_{\text{ex}} = ed$ is the dipole moments of interlayer exciton, $d = 0.6$ nm is the separation between $WSe_2$ and $WS_2$ layers and $\lambda_m = 8$ nm is the moiré periodicity. Comparing to the electron case, $U$ and $r_s$ are decreased by a factor of $\left(\frac{\lambda_m}{d}\right)^2 \sim 100$, resulting in $r_s <\sim 1$, well below the GWC threshold (*38, 39*). This can also be understood from an effective exciton charge of $\left(\frac{d}{\lambda_m}\right)e \sim 0.1e$ due to the cancellation between the electron and hole. Similarly, the Mott physics is heavily favored by the small effective charge and the corresponding enhancement of effective Bohr radius $a_*$ (*38, 39*), which gives rise to $n_0^{1/2} a_* \sim O(1)$. We thereby expect the system to be more likely a Mott insulator instead of a GWC.

6. Flavor-resolved measurements: Although the incompressible state at one interlayer exciton per moiré unit cell ($v_{ex}=1$) is necessarily a bosonic correlated insulator, its exact nature also depends on the comparison between $U$ and $\Delta E$, where $\Delta E$ is the gap between the first and second single particle bands/orbitals: the next available state after $v_{ex}=1$ is either the upper Hubbard band if $U<\Delta E$, or a different orbital if $U>\Delta E$. These two cases have been discussed for electrons in moiré superlattices(*3, 40, 49*), corresponding to a Mott insulator and a charge transfer insulator, respectively. Distinguishing the two scenarios for electrons remains an experimental challenge and is still under debate(*3, 40, 49*).

We have performed additional flavor-resolved measurements to unambiguously exclude charge transfer insulator as the nature of the observed correlated bosonic insulator. The *K* and *K'* interlayer excitons in TMDC selectively couple to one chirality of circularly polarized (CP) light(*50*). We use a linearly polarized pump light to create equal density of *K* and *K'* excitons and use a weak CP probe light to selectively create extra *K* excitons. The response from *K* and *K'* excitons are then separately recorded from the left-CP and right-CP PL. Fig. S8 A and B shows the *K* and *K'* exciton responses, respectively, depending on the pump intensity. At low exciton density, the responses from *K* and *K'* excitons are qualitatively similar (Fig. S8C). At $v_{ex}>1$, in contrast, the *K* and *K'* responses differ dramatically. Most notably, *K'* excitons show a negative response at peak I (the emission from singly-occupied exciton site), indicating that injecting extra *K* excitons will reduce the number of singly-occupied *K'* sites (Fig. S8D). This can be naturally understood from the Mott insulator picture, as an injected *K* excitons will "consume" an existing *K'* site to form a doublon site(*51*). On the other hand, the charge transfer insulator picture cannot explain our observation, since the extra *K* excitons would occupy a different orbital and should not affect the number of singly occupied *K'* sites. Furthermore, we also find that *K* and *K'* show identical responses for peak II (emission from doublon sites). This again supports the Mott insulator picture, as on a doublon site *K* and *K'* are on an equal footing and should give exactly the same emission. Because extra *K* excitons always "consume" existing *K'* sites to form doublons while doublons decay into single *K* and *K'* sites with equal probability, the system will have more *K* single sites and less *K'* single sites after doublons decay. We therefore expect to see positive *K* response and negative *K'* response of peak I with equal amplitude, which is exactly what we have observed (Fig. S8D).

Our flavor-resolved results cannot be explained by valley depolarization dynamics. This can be directly seen by considering the PL helicity

$$\eta = \frac{I_K - I_{K'}}{I_K + I_{K'}}$$

where $I_K$ and $I_{K'}$ are the intensity of PL emission from $K$ and $K'$ excitons under $K$ excitation, respectively. When there is no valley depolarization at all, all excited excitons stay in the original $K$ valley and $I_{K'} = 0$, and therefore $\eta = 1$. In the other limit of extremely efficient valley depolarization, two valleys become immediately balanced after excitation and $\eta = 0$. From these two extreme cases we can see that in the single particle picture $|\eta| \leq 1$ since both $I_K, I_{K'} > 0$. The single particle picture can explain our observation at low exciton density (Fig. S8C). We observe a rather small $\eta$, indicating efficient valley depolarization. On the other hand, at $v_{ex} > 1$ we find $I_{K'}$ becomes negative and $I_{K'} \sim -I_K$ (Fig. S8D), corresponding to $|\eta| \to \infty$. Clearly, such behavior is beyond the single particle picture and cannot be explained by valley dynamics. Instead, it is the unique consequence of a bosonic Mott insulator.

7. Estimation of exciton density at charge neutrality: The exciton density under optical pumping can be estimated through the continuous blueshift of emission energy (Fig. 1D) from exciton dipole-dipole interaction. Since interlayer excitons in WSe$_2$/WS$_2$ heterostructure have a permanent dipole moment $p_{ex}$ perpendicular to the sample plane (Fig. 1A inset), they will couple to any out-of-plane electric field $E_{op}$ and show an energy shift

$$\Delta = -p_{ex}E_{op}$$

At an interlayer exciton density of $n_{ex}$, the WSe$_2$ and WS$_2$ layers accumulate a hole and electron density of $n_{ex}$, respectively, which effectively generates an out-of-plane electrical field of $E_{dipole} = -\frac{en_{ex}}{\varepsilon_r\varepsilon_0}$ in between. Here $\varepsilon_r$ is the out-of-plane dielectric constant of the heterobilayer and $\varepsilon_0$ is vacuum permittivity. From this picture, the blueshifts of exciton energy from dipole-dipole interaction is $\Delta_{dipole} = -p_{ex}E_{dipole} = \frac{en_{ex}}{\varepsilon_r\varepsilon_0}p_{ex}$ (Ref. 36). On the other hand, when using a single gate to dope the WSe$_2$/WS$_2$ bilayer with electron density $n_e$, the electric field at the sample is $|E_{Stark}| = \frac{en_e}{\varepsilon_r\varepsilon_0}$, resulting in a Stark shift of $|\Delta_{Stark}| = \frac{en_e}{\varepsilon_r\varepsilon_0}p_{ex}$ (Ref. 36). We can therefore estimate $n_{ex}$ by comparing $\Delta_{dipole}$ and $\Delta_{Stark}$ following

$$\frac{n_{ex}}{n_e} = \frac{\Delta_{dipole}}{|\Delta_{Stark}|}.$$

At charge neutrality, the jump in exciton energy happens at pump intensity around 0.7μW/μm² (Fig. 1D), where the blueshift of peak I is $\Delta_{\text{dipole}} \approx 7$ meV compared to zero pump intensity (Fig. S13A). Meanwhile, using the bottom gate only, the exciton emission shows Stark shift $|\Delta_{\text{Stark}}| \approx 13$ meV from $v_e = 0$ to $v_e = 1$ (Fig. S13B), corresponding to an electron density $n_e = n_0$. $n_0 \approx 2.1 \times 10^{12} \text{cm}^{-2}$ is the density of one moiré filling for 1° twist(*3, 4*). We thereby estimate the incompressible state of interlayer exciton to emerge at $n_{\text{ex}} \approx 0.5 n_0 \sim 10^{12} \text{cm}^{-2}$. The above picture assumes a homogeneous sheet of charges/excitons and neglects any short-range correlation effects, therefore only offering an order-of-magnitude estimation. Nevertheless, the estimated exciton density of $\sim 10^{12} \text{cm}^{-2}$ is consistent with the bosonic correlated (Mott) insulator picture.

Another piece of evidence of $v_{\text{ex}} = 1$ at the incompressible exciton state is its smooth connection into the $v_e = 1$ electron correlated insulator upon electron-doping (Fig.2A): the electron lattice at $v_e = 1$ can only smoothly connect into a lattice of the same geometry; otherwise, the lattice is necessarily destroyed in between and the incompressibility is lost. The entire path should therefore correspond to one particle per moiré site. The incompressible state of interlayer exciton – one end of the path – then corresponds to one exciton per moiré site, i.e., $v_{\text{ex}} = 1$.

8. Calibration of exciton density with time resolved measurements: We also precisely calibrate the exciton density and filling at each pump intensity through time-resolved PL measurements. This is done in two steps. We first perform time-resolved PL (TRPL) measurement using the CW pump light as excitation light (Fig. S14A). PL emission rate is a constant over time, as expected from CW excitation. This allows us to determine the emission rate at each pump intensity (Fig. S14D). Next, we establish the relation between emission rate and exciton density by replacing the CW pump with a pulsed pump light (300 ps pulse duration, 1 MHz repetition rate) with the same wavelength (660nm) and beam profile. All other experimental configurations are also kept identical. Fig. S14B shows the time-resolved PL using the pulsed excitation light of different pulse fluences. The decay dynamics changes with pulse fluence, but is always much slower than the instrumental response function (IRF, Fig. S14C). Therefore, the emission rate immediately after time-zero corresponds to the exciton density created by the pulsed excitation light without any relaxation, which can be directly obtained from the pulse fluence.

The above procedure allows us to reliably determine exciton density without the complication from exciton lifetime or relaxation. Since the system reaches quasi-equilibrium in a short time (<1ps)(*52–54*), each measured emission rate uniquely corresponds to one exciton density at quasi-equilibrium, whether the excitation light is CW or pulsed. For example, at charge neutrality we identify $v_{\text{ex}} = 1$ at pump intensity of 0.514 μW/μm² in device D2, which corresponds to an emission rate of 595 (Fig. S14D). The same emission rate is achieved by the pulsed pump light with fluence $F = 0.26$ J/m² immediately after time zero (Fig. S14E). The exciton density is directly obtained from the pulse fluence though $n_{\text{ex}} = \alpha F/(h\nu) =$

$(2.0 \pm 0.2) \times 10^{12}$ cm$^{-2}$, where $F$ is the pulse fluence, $\alpha=(0.023 \pm 0.002)$ is absorption of WSe$_2$ at 660nm using its dielectric function(*55*) and considering the multi-layer structure of our device, $h\nu$ is the photon energy of 660nm light. $n_{ex}$ matches well with the expected exciton density $n_0 = \frac{2}{\sqrt{3}a_M^2} = 2.1 \times 10^{12}$ cm$^{-2}$ at $\nu_{ex}=1$, where $a_M \sim 7.4$ nm is the moiré periodicity considering 4% lattice mismatch and 1 degree twist angle.

We calibrate the exciton density at all pump intensity following the above procedure and obtain a pump probe PL spectrum at charge neutrality using exciton density as the vertical axis (Fig. S14F). Peak I energy shows a linear increase with exciton density, consistent with the expectation from dipolar shifts(*27, 35, 36*). The good match between two independent methods of exciton density estimation further confirms the reliability of both. By repeating the procedure at each electron density, we independently determine $n_{ex}$ across the phase diagram in Fig. 4A of the main text.

9. Inhomogeneous broadening in exciton density: an incompressible state ideally leads to a step function in chemical potential with respect to density. In practice, however, the step always has finite width due to inhomogeneous broadening. For example, broadening is universally observed in electrical capacitance measurements due to inhomogeneity in charge density across the device(*33, 56, 57*). Similarly, our pump-probe measurements collect signals from the entire probe-covered region, where spatial inhomogeneity in exciton density will broaden the transition of exciton energy. On the other hand, the inhomogeneous broadening of excitons is expected to be much larger since the exciton density depends on lifetime, which can vary by orders of magnitude depending on defects and strains(*58*). The electron density, in contrast, mainly depends on local dielectric environment and typically varies by only a few percent. This is consistent with our observation that the transition in $\Delta E_{ex}$ is sharper at $\nu_e = 1$ than at $\nu_{ex} = 1$.

10. Data analysis and fitting: The PL spectra were fitted using a two-Lorentzian model. We define exciton energy change $\Delta E_{ex} = (E_2 I_2 + E_1 I_1)/(I_1 + I_2) - E_1 = (E_2 - E_1)I_2/(I_1 + I_2)$. By taking the energy difference, we remove energy shifts from long-range dipolar interaction between interlayer excitons, which are universally reported to induce a continuous blueshift with increasing exciton density regardless of moiré effects(*35, 36*). Examples of fitting curves are shown in Fig. S15. At high pump power and/or high gate, peak II dominates and peak I largely disappears, making the fitted $\Delta E_{ex}$ unstable. In the phase diagram we only included regions where $\Delta E_{ex}$ can be reliably fit. To determine the electron one filling from absorption spectrum, we took first order derivative of reflection contrast with respect to gate voltage and focused on the intralayer moiré exciton peak around 1.76 eV (Fig. 2D), which shows a well-defined splitting/kink at one electron filling (Fig. 1E and Fig. 2B).

11. Two-component Hubbard model: Our experimental results are well captured by a two-component Hubbard model with both fermionic and bosonic species(26)

$$H = H_{ex} + H_e + U_{ex-e} \sum_{i,\sigma} n_{ex,i} n_{e,i,\sigma}$$

$$H_{ex} = -t_{ex} \sum_{\langle i,j \rangle} b_i^+ b_j + \frac{U_{ex-ex}}{2} \sum_i n_{ex,i}(n_{ex,i} - 1)$$

$$H_e = -t_e \sum_{\langle i,j \rangle, \sigma} c_{i,\sigma}^+ c_{j,\sigma} + U_{e-e} \sum_{i,\sigma \neq \sigma'} n_{e,i,\sigma} n_{e,i,\sigma'}$$

Where $\langle i, j \rangle$ represents the summation over nearest-neighbors and $\sigma$ denotes electron spin. $b_i^+$ ($c_{i,\sigma}^+$) and $b_j$ ($c_{j,\sigma}$) are creation and annihilation operators for excitons and electrons respectively. $n_{ex,i}$ and $n_{e,i,\sigma}$ are number operators for excitons and electrons. $U_{ex-ex}$, $U_{e-e}$ and $U_{e-ex}$ denotes on-site repulsive energies between excitons, electrons and exciton-electron. $t_{ex}$ and $t_e$ are the nearest-neighbor-hopping energy.

Fig. S16A summarizes the predicted exciton chemical potential $\Delta E_{ex}$ in the large $U$ limit, which matches well with the experiment (Fig. 4A). Along $v_{ex} = 0$ and $v_e \neq 0$, our model provides a natural explanation to the widely observed yet not fully understood PL doping dependence in WSe$_2$/WS$_2$(6, 8). At $v_e > 1$, the PL emission comes from a composite particle of electron and exciton on the same site, which may be considered as a correlated version of trion. However, owing to the strong correlation in the present system, it is fundamentally different from a conventional trion that has attractive interaction between the exciton and electron(20, 21, 59). The phase diagram along the vertical axis ($v_e = 0$, $v_{ex} \neq 0$) can be similarly understood. PL emission above $v_{ex} = 1$ originates from a doublon site composed of two excitons on the same site, i.e., a correlated version of biexciton. Again, it is fundamentally different from biexcitons in systems without strong correlation, which would appear at a lower energy compared to single excitons(60, 61).

Along the line of $v_{tot}=v_{ex} + v_e=1$, the jump of exciton energy with $v_{ex}$ indicates that it is generally exciton incompressible until $v_e = 1$. Because here excitons and electrons are on an equal footing, we expect the $v_{tot} = 1$ line to be also charge incompressible. This mixed correlated insulator state can be considered as a mixture of fermionic and bosonic correlated insulators, and becomes less stable than both components at higher temperature (Fig. 3 A-C). As a potential mechanism, the mixed correlated insulating state involves total electron density of $v_{tot} = 1$ and hole density of $v_h = v_{ex} <1$. While electrons have occupied all available sites, holes are at a general filling and may move around without energy penalty. At elevated temperature, holes can become mobile enough to drag electrons away and destabilize the mixed correlated insulating state.

The short-range exciton-electron repulsion $U_{\text{e-ex}}$ is a unique phenomenon in TMDC moiré superlattices with flat-bands and strong correlation, as opposed to exciton-electron attraction in e.g. monolayer TMDC. Intuitively, this can be understood from two extreme limits. In the limit of free exciton (such as TMDC monolayer without moiré superlattice), the exciton-electron interaction can be approximately understood from $-\boldsymbol{D} \cdot \boldsymbol{E}$, where $\boldsymbol{D}$ is the dipole moment of the exciton and $\boldsymbol{E}$ is the electric field generated by the electron. The exciton can always rearrange the spatial distribution of electron and hole within and align its dipole moment with $\boldsymbol{E}$, thereby leading to an attractive interaction. In the flat band limit (or strong moiré potential limit), on the other hand, the two electrons are both tightly trapped near the same local minimum of moiré potential, resulting in a strong on-site Coulomb repulsion between them. The electron-hole interaction is weaker as the hole is in the other layer. In other words, in the flatband limit the exciton and electron lose the freedom to adjust charge distribution due to the strong moiré trapping potential. Consequently, the electron-electron interaction dominates and lead to an overall short-range repulsive interaction $U_{\text{e-ex}}$. The same picture applies to short-range exciton-exciton repulsion $U_{\text{ex-ex}}$.

Because $U_{\text{ex-ex}}$ and $U_{\text{e-ex}}$ are dominated by short-range interactions, we expect contribution between two particles of the same layer to be significantly stronger than those of different layers. As a result, $U_{\text{ex-ex}}$ and $U_{\text{e-ex}}$ are expected to be in the same order as $U_{\text{e-e}}$. $U_{\text{e-e}}$ has been both measured experimentally and predicted theoretically to be in the order of 20~100meV(*4, 62, 63*). It can also be simply estimated from $e^2/\varepsilon a_M \sim 50$meV, where $\varepsilon \sim 4$ is the dielectric constant of hBN and $a_M \sim 8$nm is the moiré periodicity and also the characteristic size of wavefunction localization. Our observations of $U_{\text{e-ex}} \sim 35$meV and $U_{\text{ex-ex}} \sim 15$meV are indeed comparable to $U_{\text{e-e}}$, therefore consistent with both previous experimental results and theoretical expectations.

More quantitative estimation of $U_{\text{ex-ex}}$ and $U_{\text{e-ex}}$ e.g., from first-principle calculation involves significant theoretical challenges as the complexity scales exponentially with the number of particles. A recent theoretical report(*45*) marks the first attempt to compute excitons in a moiré superlattice using ab initio GW-BSE. Computing exciton-exciton interaction requires another major theoretical advance, and is beyond the scope of the present work. Microscopic understanding of exchange interaction between moiré excitons is an interesting and important topic, and our results provide valuable reference for future theoretical studies.

12. Discussions on compressibility measurements: In ultra-cold atom and quantum dot systems, local compressibility measurement is an established procedure to test a bosonic insulating state(*27*). Such measurement requires obtaining site-resolved particle population fluctuations, which pairs well with these bottom-up systems due to both the relatively small system size and the large site separation. On the other hand, this procedure becomes inapplicable to condensed matter systems owing to key differences between the systems: the much larger number of particles and much smaller site separation in top-down systems make it impractical to resolve individual sites. On the other hand, because condensed matter systems are typically in the thermodynamic limit, the compressibility $\kappa$ can be directly obtained from its original definition $1/\kappa = n^2 d\mu/dn$ without involving fluctuation-dissipation theorem, where $n$ is the particle density and $\mu$ is its chemical potential. Therefore, the well-established procedure to measure compressibility in condensed matter physics community is through thermodynamic quantities, such as electrical capacitance.

In this sense, our technique carries forward a well-established procedure in condensed matter physics community and applies it to a novel context of optical measurement and excitons. Conventional optical measurements such as PL cannot obtain compressibility, since they collect responses from all excitons in the system. This is similar to the charge case where electron chemical potential can be obtained by examining electrons at the Fermi level, but it is obscured if the measurement collects responses from all electrons in the system. Meanwhile, site-resolved measurement is technically too challenging given the small (<10nm) site separation. Our pump probe technique offers the unique capability to isolate low energy excitations of the bosonic ground state and determine exciton chemical potential, thereby providing a powerful toolbox to investigate novel correlated states of bosons in condensed matter systems.

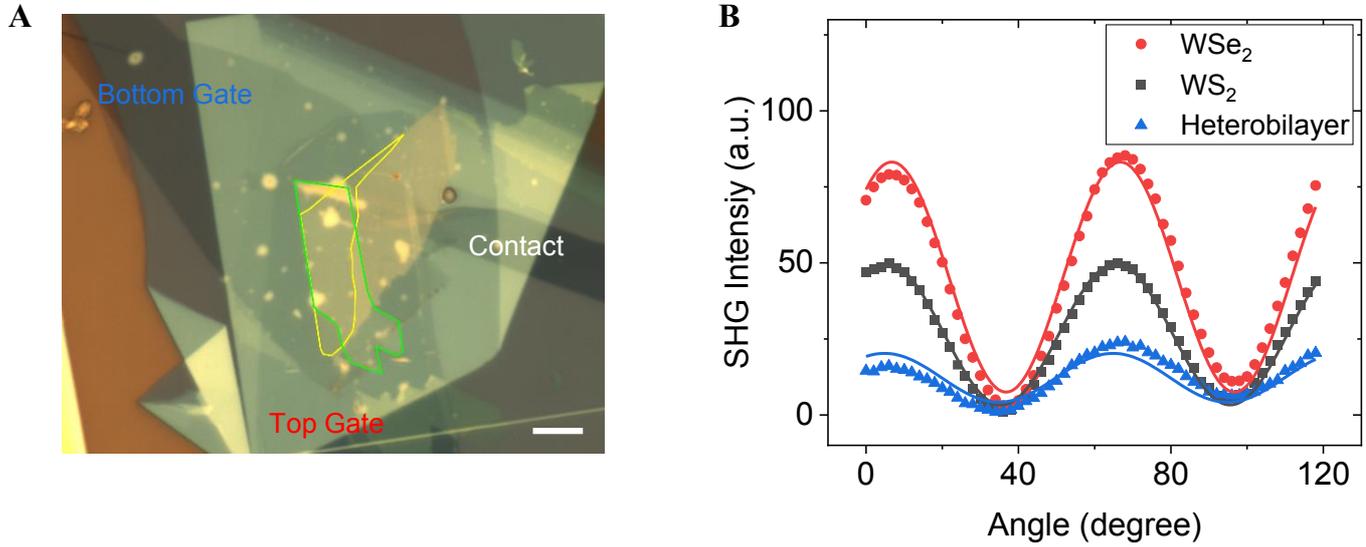

**Fig. S1. Optical image and SHG measurement of representative devices. A,** optical image of a typical dual-gated 60-degree aligned WSe$_2$/WS$_2$ device D1. Yellow and green solid lines denote the sample edges of monolayer WSe$_2$ and WS$_2$, respectively. The scalar bar is 5 μm. **B,** SHG measurement on monolayer WSe$_2$ (red circles), monolayer WS$_2$ (grey squares) and heterobilayer (blue triangles) region. Solid curves are fits to the data using A + Bcos2[3(φ – φ0)], where φ is the excitation polarization angle, φ0 is the crystal orientation and A and B are fitting parameters. For WSe$_2$ and WS$_2$ monolayer, the crystal orientation φ0 is 6.7° ± 0.2° and 5.6° ± 0.1° respectively. The twist angle is determined to be 1.1° ± 0.3°. Since the SHG signal from the bilayer region is much weaker than that from the monolayers, we conclude that the WSe$_2$ and WS$_2$ are nearly 60-degree aligned.

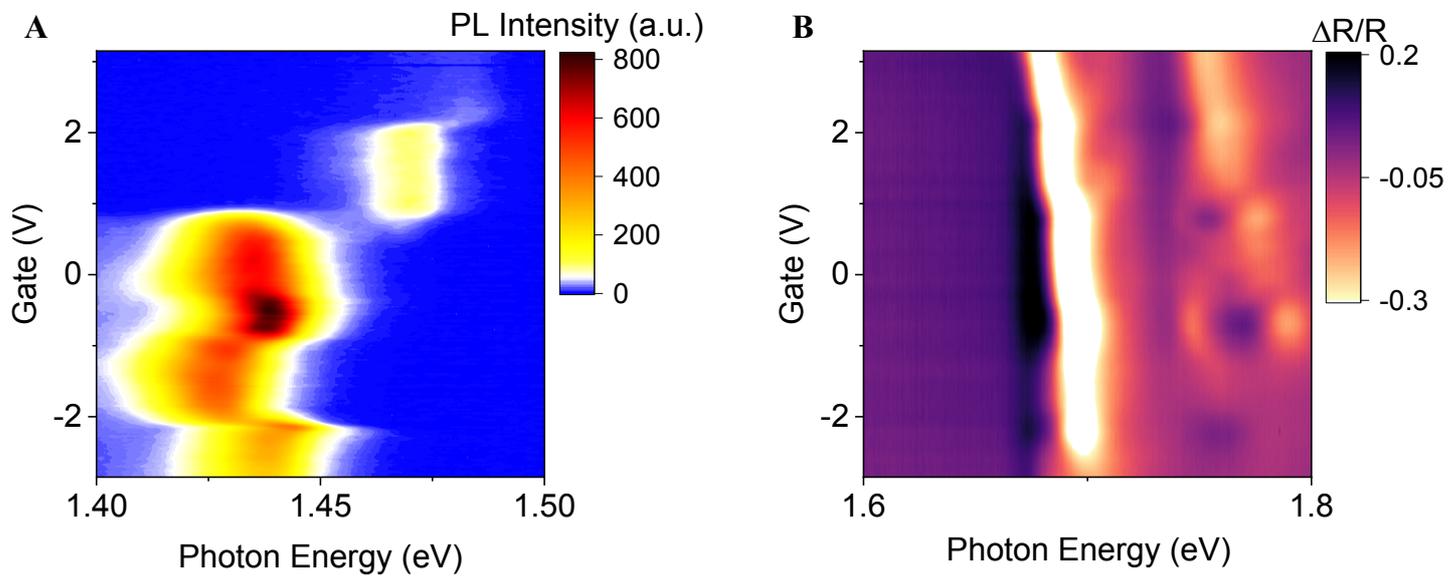

**Fig. S2. Complete PL and absorption spectra. A,B,** Doping-dependent PL (A) and absorption (B) spectrum of a 60-degree aligned $WSe_2/WS_2$ moiré bilayer device D1 at zero pump intensity, from which Fig. 1-3 in the main text is taken.

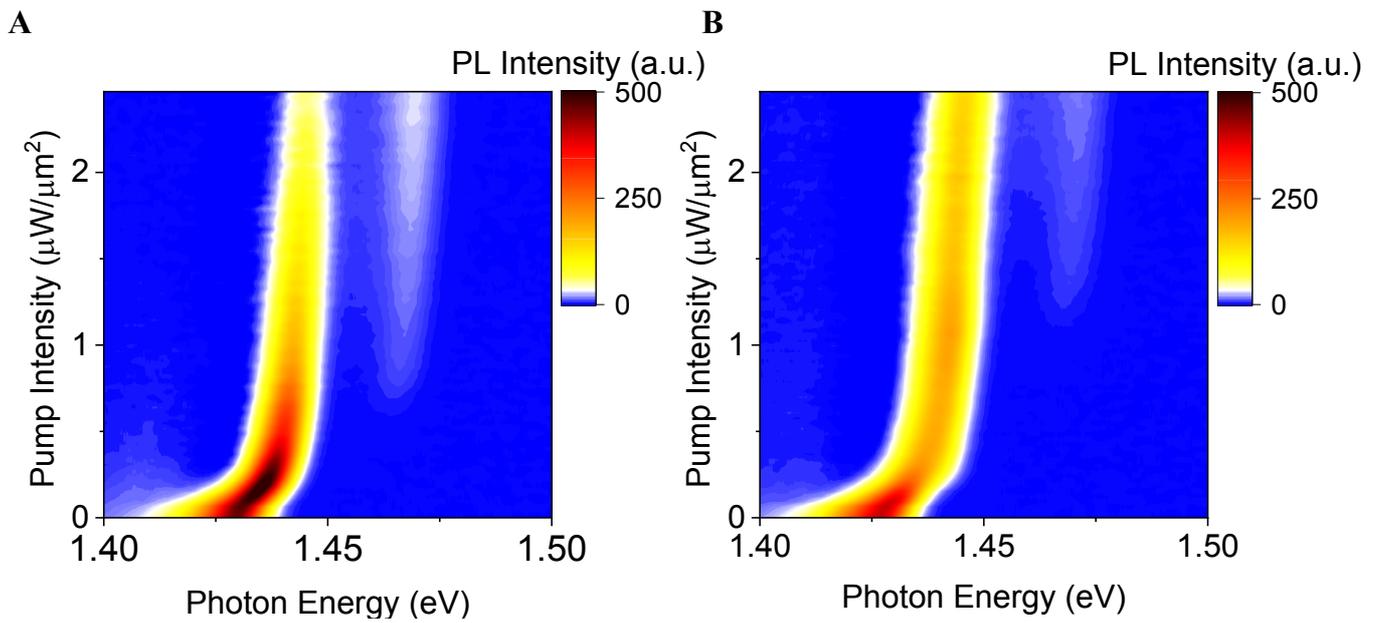

**Fig. S3. Pump probe PL spectrum on the hole doping side (device D1). A, B,** Pump probe PL spectrum at $\nu_h$=0.1 (A) and $\nu_h$=0.5 (B). Hole doping does not cooperate with excitons to form a mixed correlated insulator and instead destabilize the exciton lattice. This could possibly be due to excitons and holes residing different high symmetry points in a moiré unit cell.

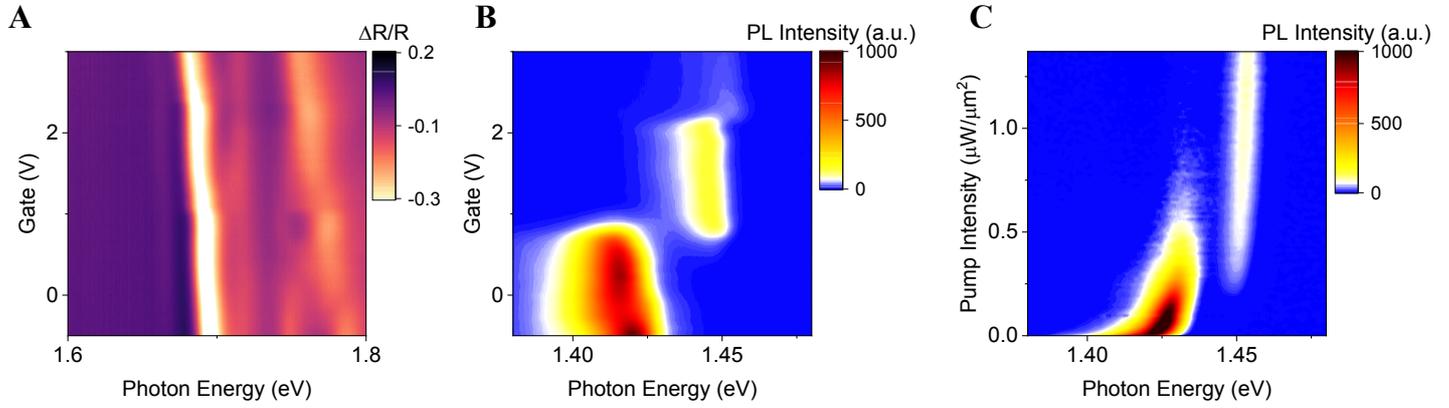

**Fig. S4. Results from another 60-degree aligned moiré bilayer. A,B,** Doping-dependent absorption (A) and PL (B) spectrum of another 60-degree aligned WSe$_2$/WS$_2$ moiré bilayer device D2. **C,** Pump intensity-dependent PL at charge neutrality. A qualitatively similar jump in interlayer exciton energy is observed, indicating emergence of a bosonic correlated insulator composed of interlayer excitons.

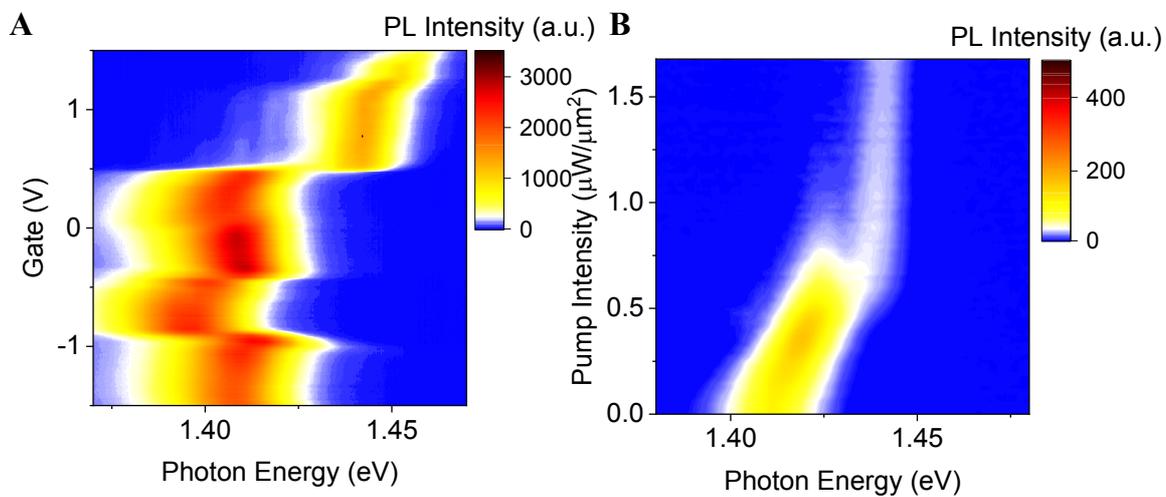

**Fig. S5 Results from a 60-degree aligned moiré bilayer with broader PL linewidth. A, B,** Gate-dependent PL at 1.65 K of a 60-degree-aligned $WSe_2/WS_2$ device D3 (A) and pump-probe PL at charge neutrality (B). The observed $U_{e\text{-}ex}$ and $U_{ex\text{-}ex}$ are consistent with devices D1 and D2.

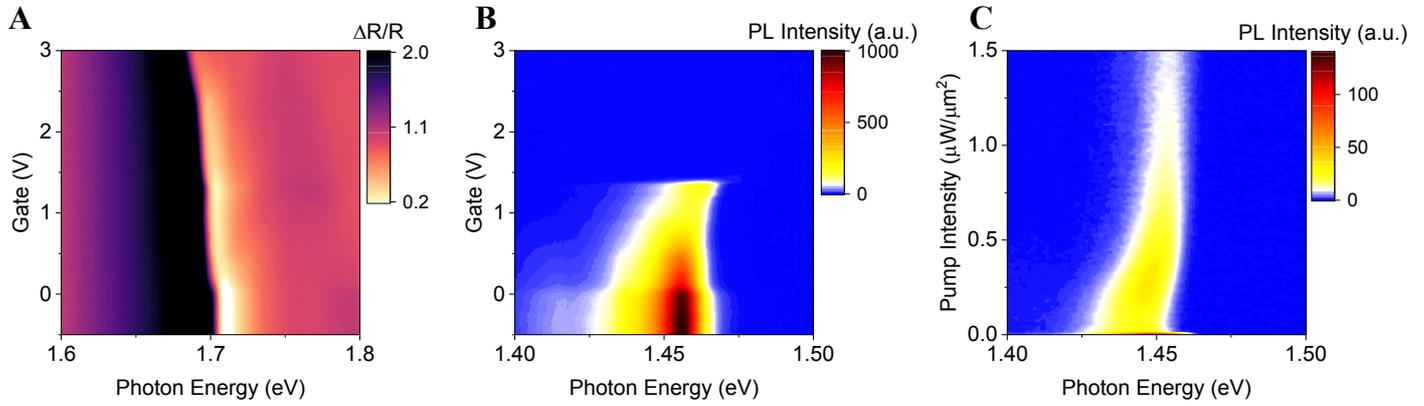

**Fig. S6. Results from a misaligned sample. A,B,** Doping-dependent PL (A) and absorption (B) spectrum of a 3-degree-misaligned $WSe_2/WS_2$ bilayer device D4. **C,** Pump intensity-dependent PL at charge neutrality. No jump in the interlayer exciton energy is observed due to the lack of strong correlation. On the other hand, the continuous PL blueshift from long-range dipolar interactions is still observed.

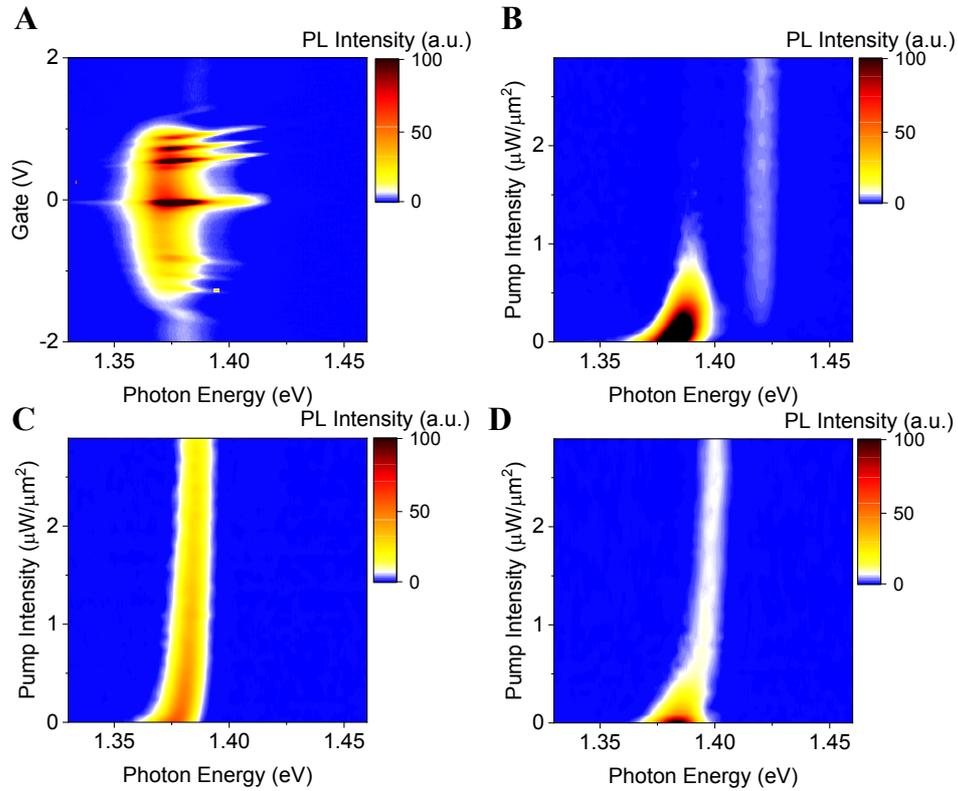

**Figure S7. Pump-probe PL spectrum of a 0-degree aligned sample D5. A,** Gate-dependent PL at 15 K. **B,** Pump-probe PL at charge neutrality, which shows a similar jump as 60-degree samples. **C, D** Pump-probe PL spectrum at $\nu_h=1/3$ (C) and $\nu_e=1/3$ (D). Both hole and electron doping do not cooperate with excitons to form a mixed correlated insulator and instead destabilize the exciton lattice. B-D are measured at base temperature of 1.65 K.

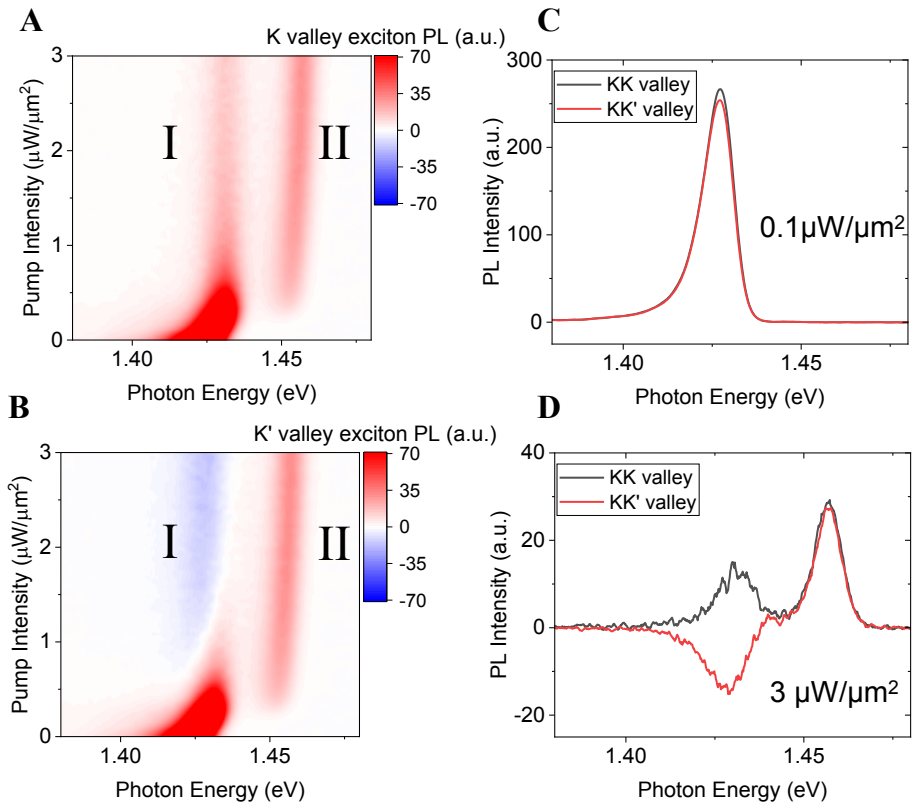

**Fig. S8. Flavor-resolved pump-probe PL of device D2 at charge neutrality. A, B,** A linear pump light is used to generate equal numbers of *K* and *K'* valley excitons, while a CP probe selectively excites extra *K* valley excitons. *K* valley (A) or *K'* valley (B) PL response induced by the probe light is collected separately. **C, D,** Linecuts of A and B at low (C) and high (D) exciton density. KK(KK') valley refers to the probe injecting *K* valley excitons and PL detecting *K*(*K'*) valley excitons. Pump light is always linear.

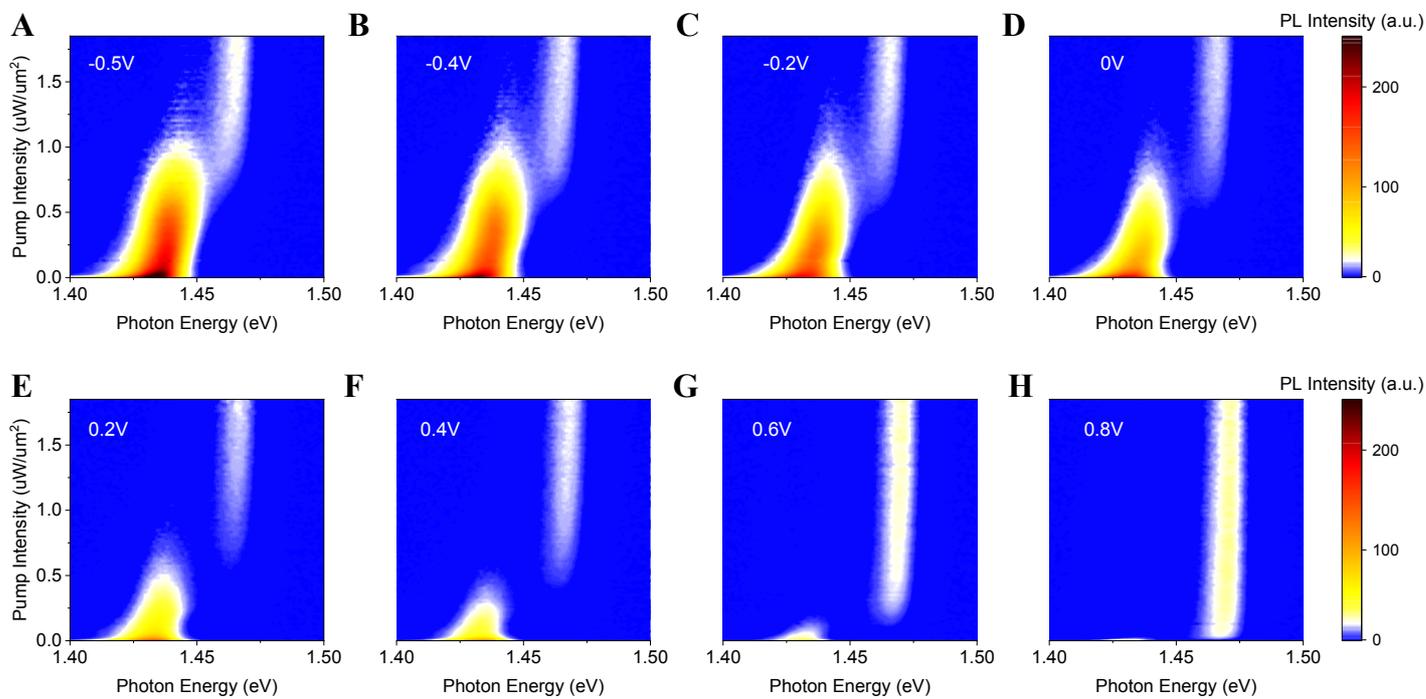

**Fig. S9. Pump intensity-dependent PL spectra at 15 K** (device D1) for gate voltages from -0.5V (charge neutrality) to 0.8V (electron-one-filling).

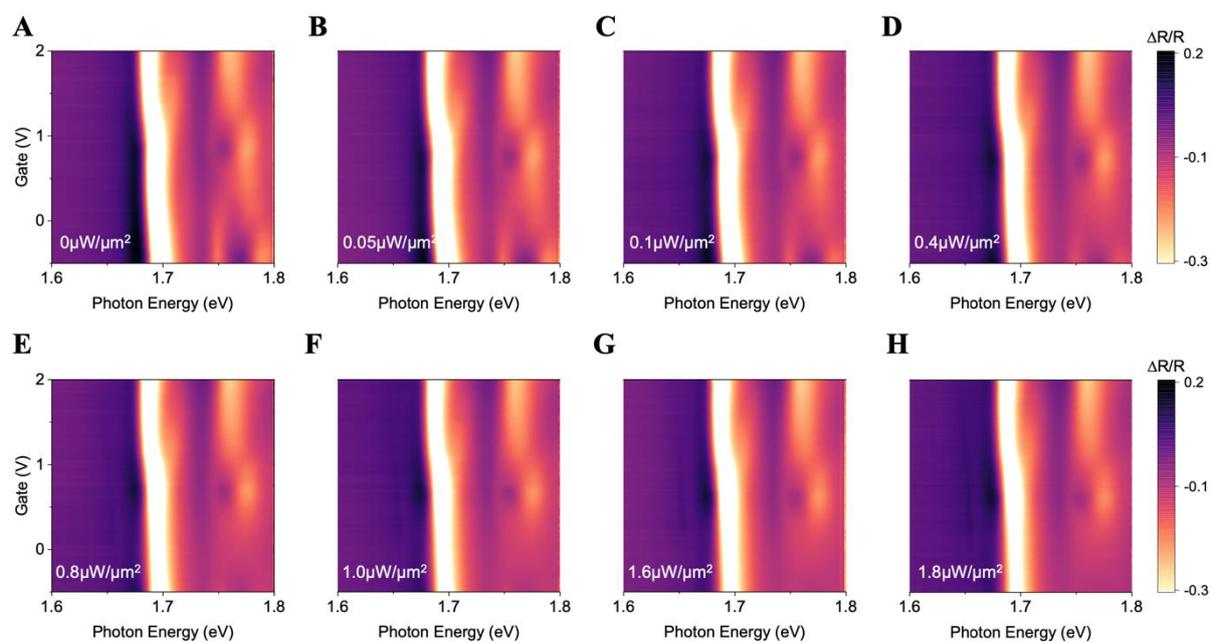

**Fig. S10. Gate-dependent absorption spectra at 15 K** (device D1) for pump intensities from 0μW/μm$^2$ to 1.8μW/μm$^2$.

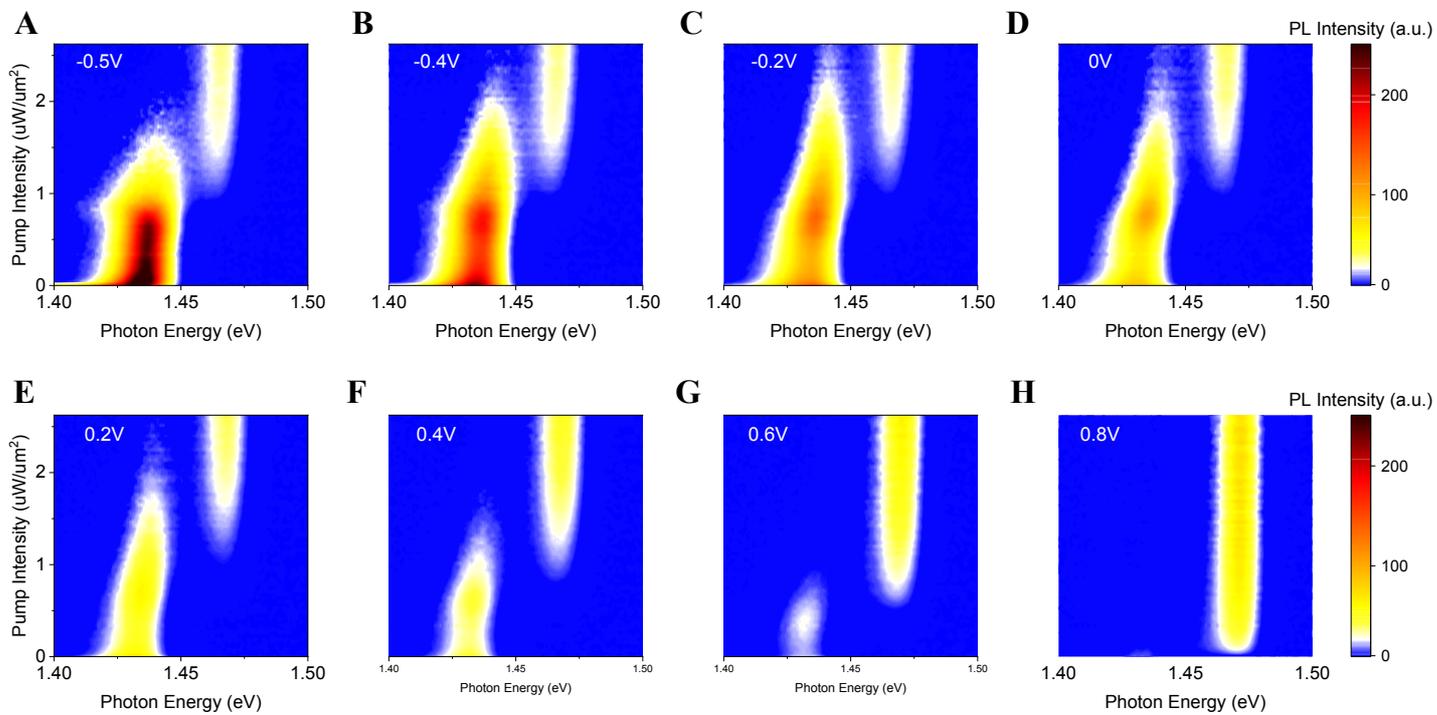

**Fig. S11. Pump intensity-dependent PL spectra at 30 K** (device D1) for gate voltages from -0.5V (charge neutrality) to 0.8V (electron-one-filling).

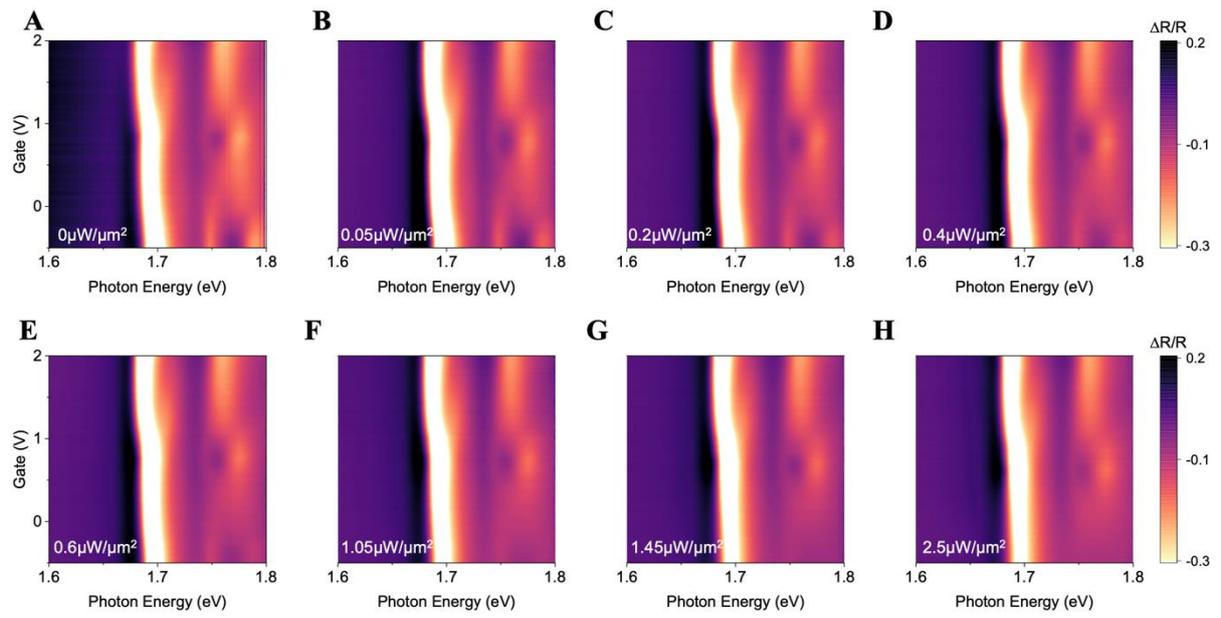

**Fig. S12. Gate-dependent absorption spectra at 30 K** (device D1) for pump intensities from 0μW/μm$^2$ to 2.5μW/μm$^2$.

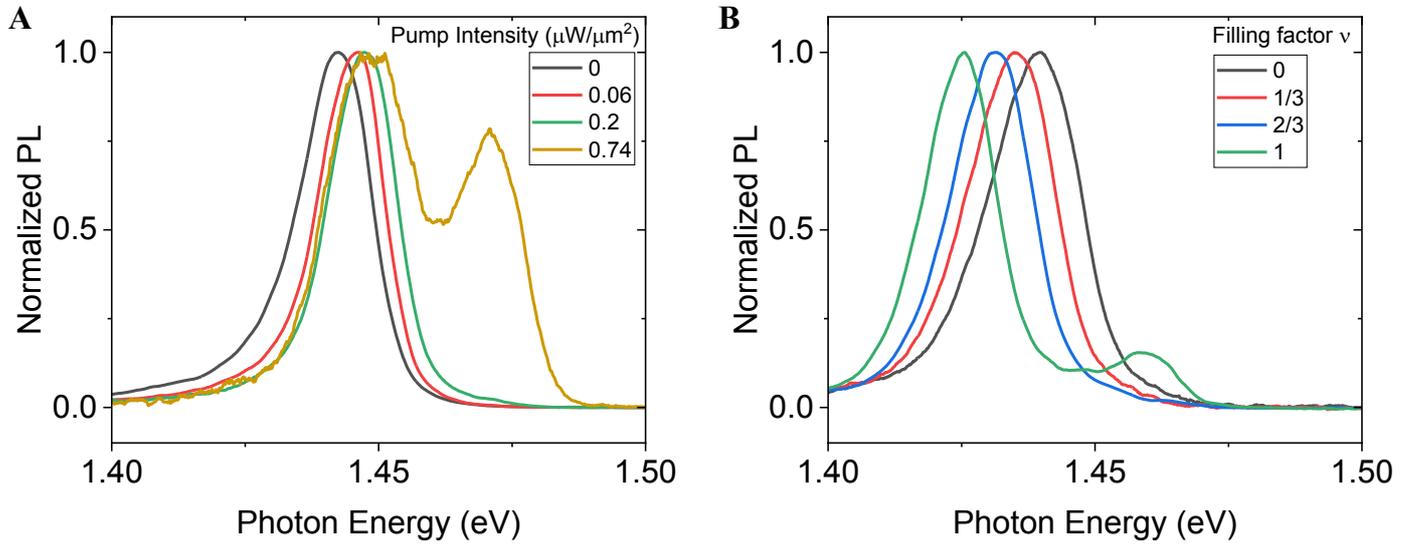

**Fig. S13. Estimation of exciton density. A,** Pump-probe PL spectrum at charge neutrality and selected pump intensity. The low energy peak (peak I) shows a continuous blueshift with pump intensity from dipolar interaction between interlayer excitons, which can be used to estimate the exciton density generated by pump. **B,** PL spectrum at zero pump intensity and selected electron fillings tuned by a single gate. Peak I shows a continuous Stark shift with respect to the filling, which can be used to calibrate the relation between carrier density and PL shift.

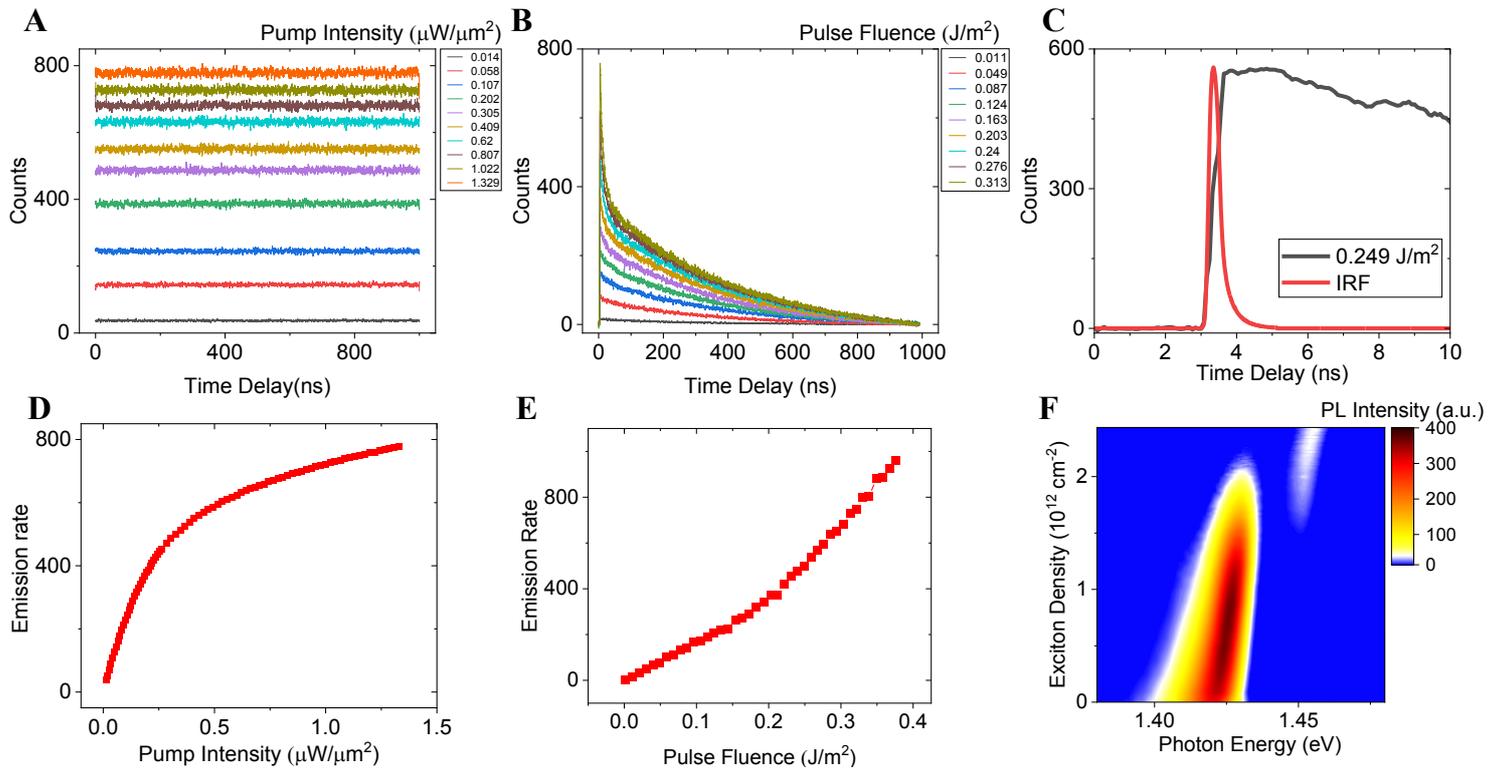

**Fig. S14. Calibration of exciton density via time resolved PL (TRPL) measurement. A,** TRPL of device D2 at charge neutrality using a 660nm CW pump light of different pump intensity. **B,** Same as A but with a 660nm pulsed pump light (300ps duration, 1MHz repetition rate). **C,** Comparison between IRF and PL dynamics indicates negligible exciton relaxation immediately after time zero. **D,** Emission rates from the CW pump light of different intensities. **E,** Emission rates from the pulsed pump light of different fluences. **F,** pump-probe PL spectrum of device D2 at charge neutrality with calibrated exciton density as the vertical axis.

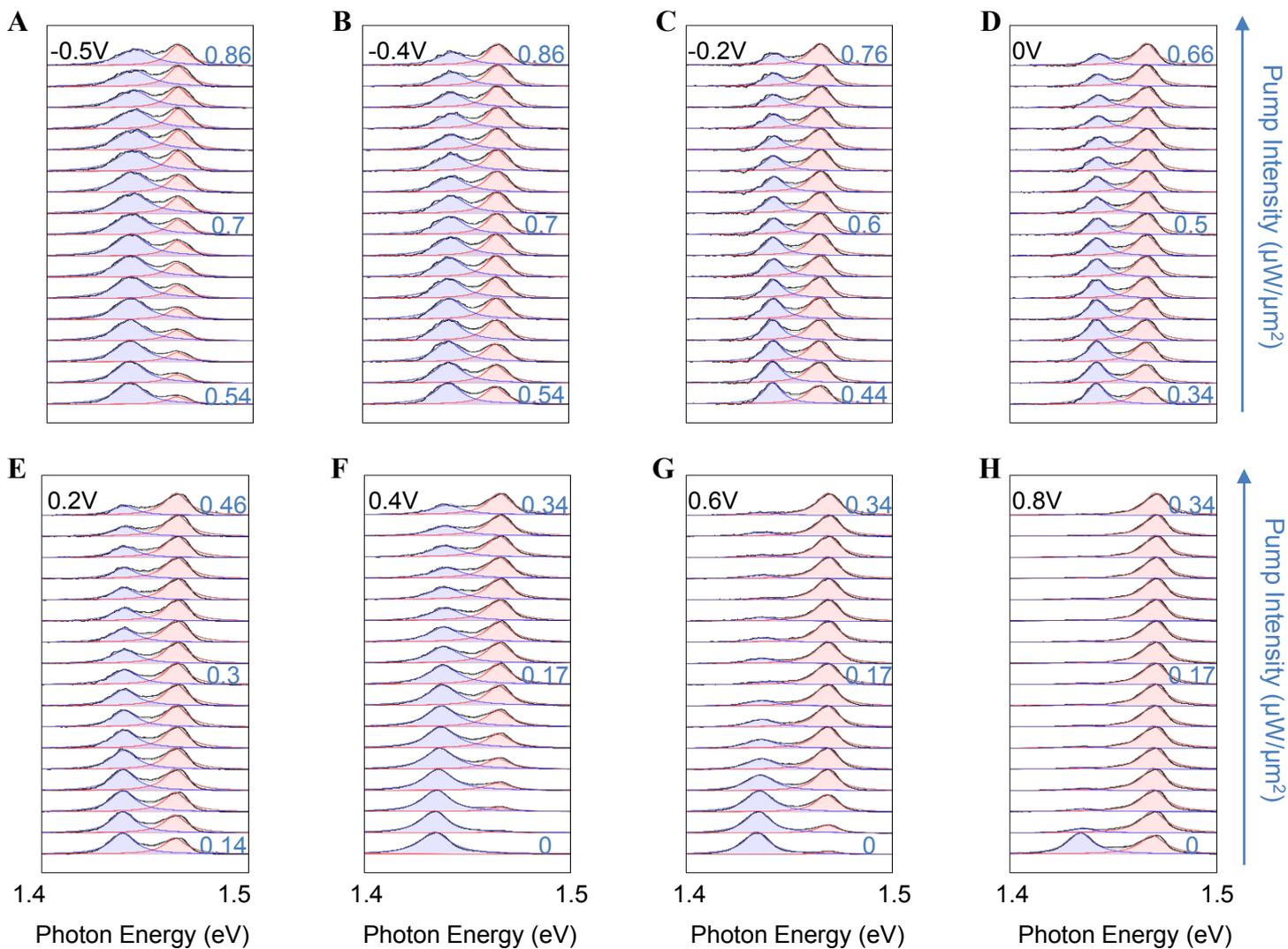

**Fig. S15. Waterfall plots near transition region. A-H,** Waterfall plots near spectral jumps of pump probe PL spectra of device D1 at 1.65 K. Black and grey curves are experimental data and two-peak Lorentz fitting, respectively. Blue and red curves are fitted peak I and II.

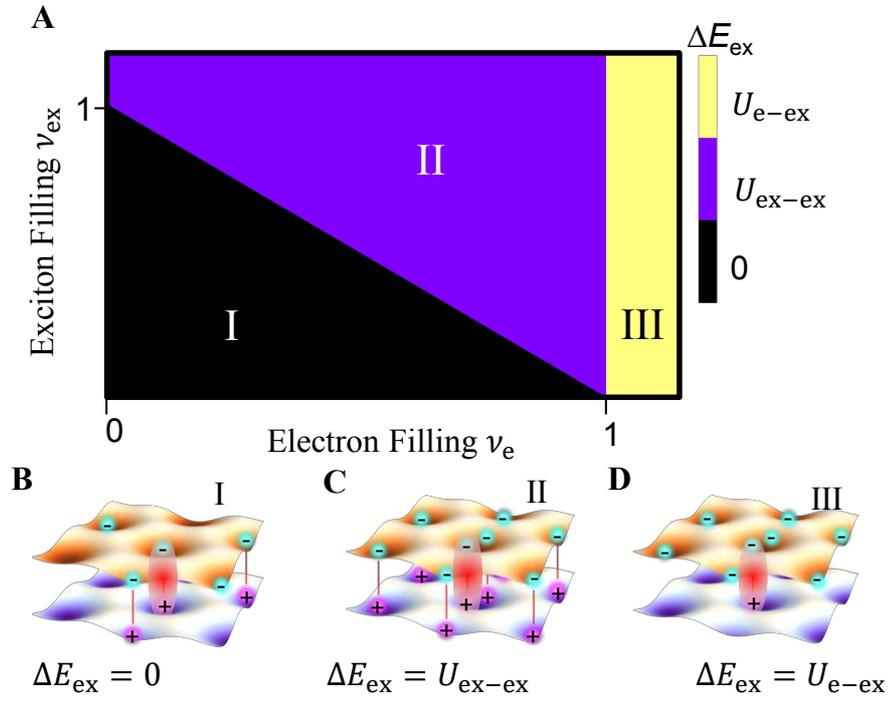

**Fig. S16. Mixed Hubbard model. A,** $\Delta E_{ex}$ with respect to $\nu_e$ and $\nu_{ex}$ predicted from a mixed Hubbard model in the large U limit. Boundaries between region I, II and II, III correspond to $\nu_{tot} = 1$ and $\nu_e = 1$. **B-D,** Schematic plots showing the energy required to add one interlayer exciton (red) into the system for region I (B), II (C) and III (D) in the phase diagram. Additional energy cost of $\Delta E_{ex} = U_{ex-ex}$ and $U_{e-ex}$ is required in region II and III, respectively, from on-site repulsion between excitons and between electron-exciton.